# A Stock Options Metaphor for Content Delivery Networks

**Elias Vathias[1] - Stathes Hadjiefthymiades[2]**

**Abstract** The concept of Stock Options (SOs) is used to address the scarcity of resources, not adequately addressed by the previous tools of our Prediction Mechanism (PM). Using a Predictive Reservation Scheme (PRS), network and disk resources are being monitored through well-established techniques (Kernel Regression Estimators) in a given time frame. Next, an Secondary Market mechanism (SMM) significantly improves the efficiency and robustness of our PRS by allowing the fast exchange of unused (remaining) resources between the Origin Servers (OSs - CDN Clients). This exchange can happen, either by implementing socially optimal practices or by allowing automatic electronic auctions at the end of the day (EOD) or at shorter time intervals. Finally, we further enhance our PM; SOs are obtained and exercised, depending on the lack of resources at the EOD. As a result, OSs may acquire resources (if required) at a normal price. The effectiveness of our mechanism further improves.

## 1. Introduction

Nowadays, the adoption of Content Delivery Networks is standard practice for many content-generating entities. The population of Origin Servers (OS) is constantly increasing with content that becomes multimedia-richer, “heavier” and, structurally, far more complex. In this evolving ecosystem, we aim at rationalizing the reservation and use of CDN resources from both the side of the client (OS) and the CDN. The CDN gradually shifts its attention to more fine-grained resource management schemes that minimize overheads (unused capacity, etc.). The use of CDN resources needs to be carefully planned in the space-time domains.

[1] E. Vathias
National and Kapodistrian University of Athens, Greece
email: evathias@di.uoa.gr

[2] S. Hadjiefthymiades
National and Kapodistrian University of Athens, Greece
email: shadj@di.uoa.gr

Schemes and techniques for more accurate resource claims are of common interest as they positively impact not only the involved clients (OS) but also the CDN operator/provider. Resource utilization can be maximized and, thus, advance the pertinent economy in total (i.e., improving the position of both sellers and buyers). CDN providers can deliver better services at more competitive prices to a much wider audience, especially in relation to other CDN providers that do not use this framework. Furthermore, improving resource utilization results in the reduction of network congestion.

We adopt an optimized framework for the methodical management of CDN resources. Specifically, we borrow concepts, schemes, and techniques from the capital market [12] [19] [20] [22]. We treat the CDN resources (assets) as (capital) stocks. The stockholders are the content-generating organizations (OS). Stocks are purchased dynamically, according to the client needs for resources, aiming to sustain the experienced visitor load efficiently.

To better follow the exact resource use and protect against resource misuse, CDN monitors the incoming traffic for each client and tries to establish load predictions for the immediate future. This knowledge is of the utmost importance to the rationalization of CDN resource management. The CDN learns the temporal distribution of resource use for each client for both bandwidth and disk space.

Load prediction allows CDN to reserve resources for the OS well beforehand and, thus, cope with the fluctuating load. Resources can be claimed from the CDN but also returned to the free pool (Fig. 1) for use by other clients. A tool for the trading of CDN resources, which improves the resilience of the primary mechanism to prediction failures, is the capital Secondary Market (SM). CDN resources are traded in the SM by the actual proprietors (i.e., clients that have already acquired resources but experience load incompatible to their expectations/predictions).

In this paper, we introduce another tool that further improves the resilience of the forecasting and resource swapping mechanisms, the capital SOs. CDN purchases SOs, for each OS, at prices and for the duration defined by models such as that of Black-Scholes (BS), Barone-Adesi & Whaley [24], Bjerksund & Stensland, Ju & Zhong [23], Binomial and Trinomial Trees, etc. These SOs, and exercised if and when they are needed (i.e., if an OS runs out of resources at the end of the billing period and cannot find any available through the SM).

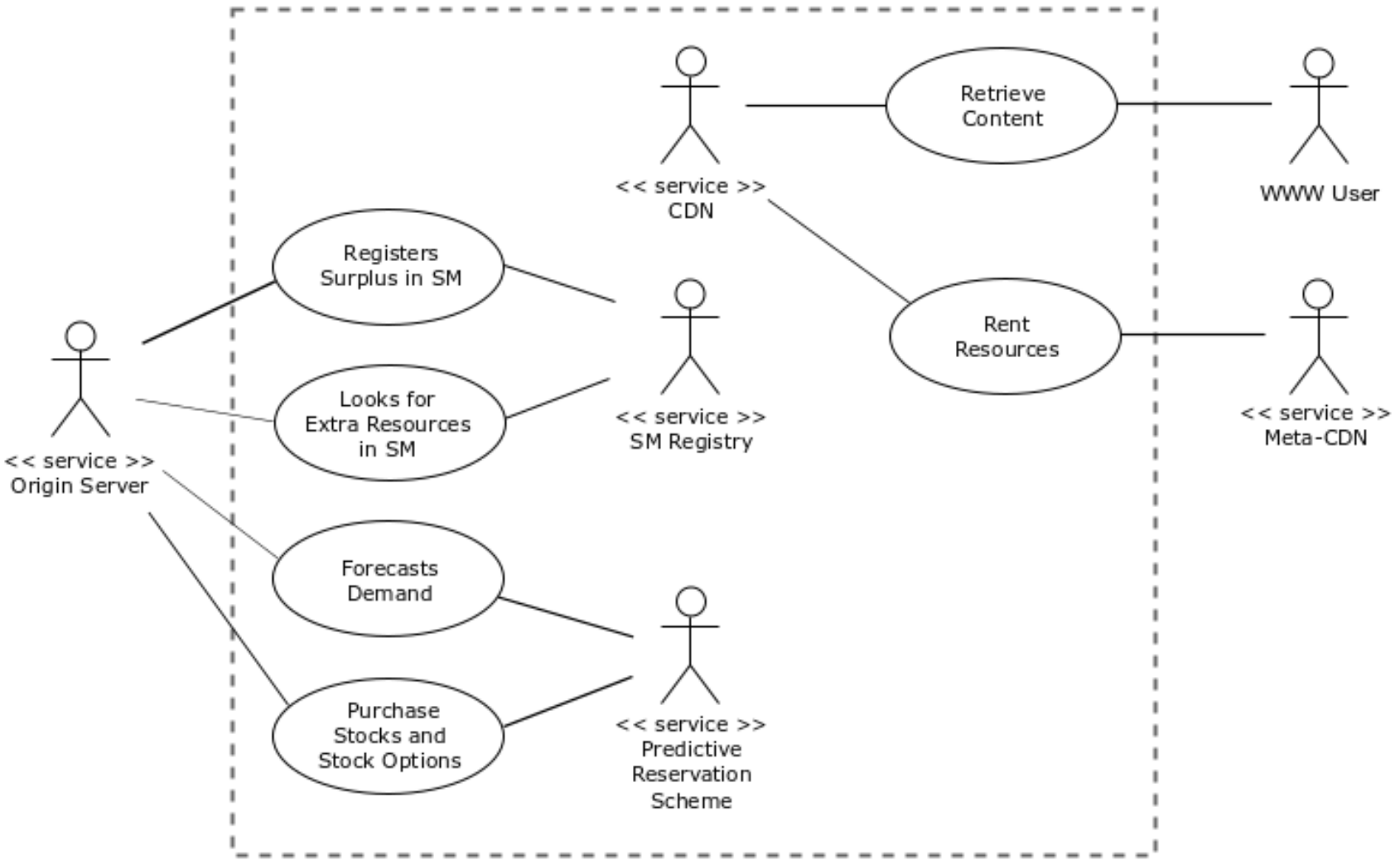

**Fig. 1.** Basic Framework Structure

This scheme is particularly important for CDN providers that “lease” resources (disk space and bandwidth) from other operators. The latter stakeholders (higher level entities) constitute a meta-level in the CDN ecosystem (meta-CDN), which allows for the implementation of the resource management scheme (Fig. 1). The traded resources are finite; hence, the CDN needs to plan its acquisition and management carefully. The proposed scheme offers this opportunity as it allows more rightly predicted resource claims but also deals with prediction failures. Therefore, all the involved stakeholders maximally exploit the resources that the meta-CDN offer.

In this paper, based on the mechanisms we presented in our previous two papers [20] [22], we are introducing another mechanism that further enhances the optimization of the use of CDN resources.

Describing the overall framework, initially, the PM evaluates traffic and introduces load forecasts for the immediate future. Then the SM mechanism allows the horizontal (across different OS) trading of resources, to address predictions failures. Finally, the new SO mechanism ensures the ability to obtain resources at a normal price, if needed, at the EOD and at the cost of acquiring the options, by observing the failures, as a whole, of the previous two mechanisms (PM and SM).

We access the performance of the discussed scheme using extensive real traces taken from high traffic OS (Chambers of Commerce, Universities, Popular Sports Content Sites, etc.). Our findings show that the introduction of the SOs enables our framework to further “absorb” prediction failures and optimize CDN resource management. We compare the efficiency of our scheme with that of other models that are currently being used. We argue that this capital market metaphor is a good candidate for structuring the modern WWW ecosystem.

The paper is organized as follows. Section 2 discusses the Related Work. Section 3 presents the architecture of our framework. Section 4 describes in detail our simulations plans as well as the metrics and results of them. Finally, Section 5 concludes the paper and discusses future work.

## 2. Related Work

Internet resources pricing is a crucial factor for the efficient allocation of Internet resources as also as the determinant of the profit [8]. CDNs have to deal with the cost of interconnection and traffic on their networks. At the same time, they are trying to increase their revenues by choosing effective pricing strategies that have to do with the efficient allocation of Internet resources.

In view of the above, there are three primary **pricing models**: flat pricing, usage-based pricing, and congestion pricing. Flat pricing [26] was sufficient for the early stage of the internet when applications were simple, and resources were sufficient. The increase of network content in conjunction with the lack of incentives for efficient network resource usage resulted in the overall network performance degradation. To address this, usage-based pricing [27,28,29] was proposed. The main idea was that if the charge were usage-based, and since Edell and Varaiya [52] showed that users are highly sensitive to pricing, a fairer and more efficient use of resources would be feasible. Among problems that had to be addressed were the privacy issues in processing audit and the charge of non-expected traffic, i.e., ads, spam, etc. For example, the 95th percentile pricing became an industry standard. In this method, the peak flow within 5% of the total time (36 hours per month) is free of charge.

However, network traffic continued to increase, exacerbating congestion. As a result, the aforementioned pricing became more complex, leading to a relatively dynamic pricing model "congestion pricing" [26, 30, 31, 32, 33, 34, 35, 36, 37, 38] which has been studied extensively. Congestion pricing dynamically sets prices that can reflect approximate real-time network resource usage and, especially when the network is busy, encourages shifting the traffic from peak time to non-peak time, reducing congestion.

Here it is worth noting that, in the CDN context, congestion reduction is not a key goal of pricing. Content providers subscribe to CDN services precisely to overcome Denial of Service Attacks or Flash Crowds. Moreover, the traffic of different content providers is unlikely to surge at the same time. So CDNs can temporarily adjust their infrastructure to handle the traffic spike and improve the availability of content. As a result, the research on congestion pricing, while relevant, does not directly affect the CDN.

With regard to the pricing of the different models applied, **pricing mechanisms** can be categorized into two types: best-effort and QoS-enabled. In best-effort type, users are charged according to access rate or resource usage. In QoS-enabled type, ISPs tend to serve different data streams with different QoS and price levels.

Generally, for best-effort networks, pricing is always done at the edge of networks and incurs a lower overhead cost, while QoS guaranteed services involve a higher audition and accounting cost. Priority-based pricing was first proposed by Cocchi et al. [39, 40] to perform service layering with the corresponding pricing. Another well-known proposal of QoS differentiation was Odlyzko's Paris Metro Pricing [41], which divides the network into subnets and charges them differently. QoS guaranteed network architectures (e.g., IntServ and DiffServ), and their corresponding pricing mechanisms have been widely studied [25, 42, 43, 44, 45, 46].

With regard to network **pricing methods**, there are two main models for the determination of the appropriate price levels: system optimization models and strategic optimization models. System optimization models are mainly based on optimization theory like the concept of Network Utility Maximization (NUM) framework proposed by Kelly [33] which is the initial work of Internet system optimization, as well as other works on network utility maximization [48]. Strategic optimization models, i.e., considering strategic behaviors of the others when setting prices or making other decisions [50, 51], are based on non-cooperative games [47, 49].

Following those previously mentioned, there have been several recent studies in various aspects of the Caching and CDN technologies, including resource management and pricing.

## 2.1 Caching

More specifically in the field of multi-level caching, it is observed that some clients, exhibiting "aggressive" behavior, tend to monopolize the cache disk space, thus, enjoying high hit rates. Other clients, at the same time, are confined to restricted disk space, and suffer the eviction of "important" objects, thus, experiencing numerous cache misses and underperformance. The authors in [1] proposed a framework that discourages monopolizing the cache disk space by a minority of clients, while rewarding clients that contribute to the overall hit rate by allowing them to use more disk space. Hosanagar et al. [53, 10] find that the adoption of traditional best effort caching will decrease as OSs move towards dynamic content and simultaneously seek accurate business intelligence regarding website usage. They argue that CDNs can play an important role intermediating between OSs that seek the benefits of edge delivery and ISPs that can install servers at the edge of the network. OSs can enjoy the benefits of edge delivery of content without incurring the costs of best effort caching.

## 2.2 Distributed Group of Nodes

Another standard model for studying Caching Proxies, CDN and P2P technologies is that of a distributed group of nodes, where each of them uses the storage capacity to create copies of objects, either through replication (permanent copies) or through caching (temporary copies) and render them available to local and remote users. In the case of replication, the authors in [14] propose a Two-Step Local

Search (k) algorithm which protects nodes from mismanagement. In the case of caching, the authors in [15] propose detection, addressing, and adjusting mismanagement mechanisms.

### 2.3 Content Delivery Networks

In the area of CDN, when several Service Classes with different qualities of service are offered to the publishers, the authors in [6] discussed a simple differentiated service type architecture for content delivery networks and proposed a pricing scheme to complement this architecture and provide fair service to the subscribed publishers. They also have investigated and suggested methods to determine the optimal pricing of these services, the optimal allocation of resources between the services and the optimal number of services to be offered.

In similar research [9], using analytical models, the same authors addressed the optimal pricing of the offered services and studied how external factors such as the cost of bandwidth and security issues affect pricing. For example, they found that (a) declining bandwidth costs will negatively impact CDN revenues and profits, (b) CDNs will have to lower prices in light of increasing security concerns associated with content distribution or they will need to invest in developing and deploying technology to alleviate the security concerns and (c) larger CDN networks can charge higher prices in equilibrium, which should strengthen any technology-based economies of scale and make it more difficult for entrants to compete against incumbent firms.

Hosanagar et al. [11] study the optimal pricing for a monopoly CDN. They find that traditional usage-based pricing plans should entail volume discounts when subscribing content providers have similar levels of traffic burstiness, but that volume discounts can prove suboptimal when traffic burstiness is highly heterogeneous. Moreover, they find that profitability from a percentile-based pricing plan can be substantially higher than traditional usage-based billing.

The authors in [5] investigated the maximization of the benefits gained by OS as well as by CDNs. Among the results of their research, they found that (a) as the CDN lowers its price, it receives higher investment from the publishers, b) lowering the price more than a specific level reduces the revenue because the CDN has a limited cache space and the publishers' requests cannot be completely satisfied, (c) the surrogate revenue is maximized when the total publisher demand is equal to the CDN cache space, d) surrogates need not be very close to the users and e) while the system optimum investment maximizes the total of publisher utilities, it reduces some of the publishers individual utilities.

### 2.4 Auctions in Resource Management Area

The authors in [13] propose an auction approach to dynamically allocate the spectrum in a secondary market, in order to enable better use of the wireless spectrum.

### 2.5 Our Contribution

Our work differs from the previous works in the sense that a) it uses load monitoring, modeling, and prediction b) it uses capital market instruments (secondary markets and stock options), based on corporate finance, c) it offers a more fine-grained resource management scheme that reduces overheads (unused capacity etc.) and enables the CDN to carefully plan its resources needs. As a result, all the involved players benefit from the rationalized use of CDN resources. Furthermore, improving resource utilization results in the reduction of network congestion.

Our previous model [20] [22] by using monitoring, modeling, prediction and exchange mechanisms gave all the involved players (Origin Servers and the CDN itself) a significantly better image for the resource needs arising from the user agent (UA) visits and, as a result, improved the rationalization of CDN resources usage.

In our current work, by extending [20] [22], we elaborate on the SO mechanism operation, simulating different plans and ways in which, based on the failures of our mechanism so far, contracts are purchased that allow the acquisition of resources at advantageous prices in the future if and when they are needed (i.e., if an OS runs out of resources at the EOD and cannot find the total amount required through the SM). Our framework efficiency further improves.

## 3. Architecture

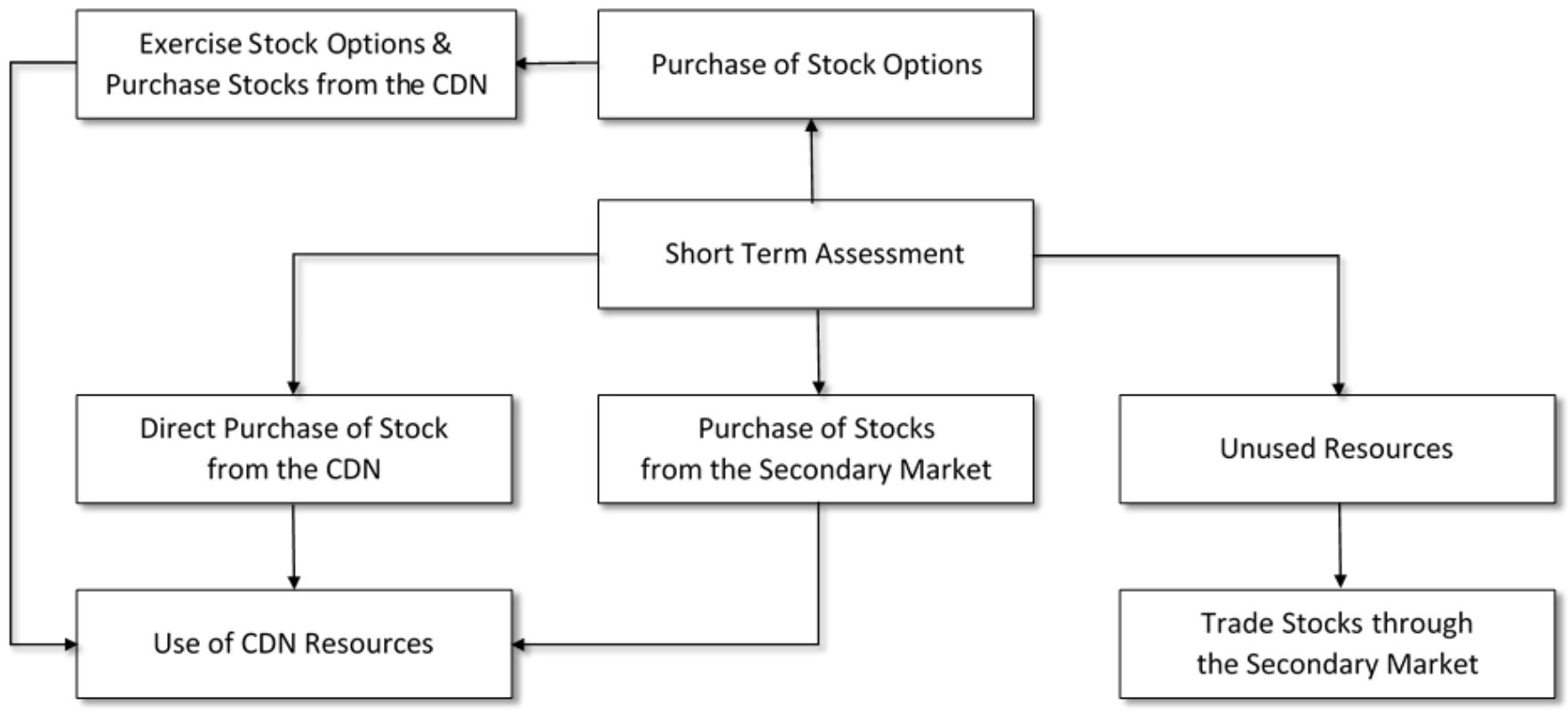

**Fig. 2.** Predictive Reservation Scheme (PRS)

In this section, we recapitulate the architecture of our mechanism so far, which is presented in detail in [20] and [22] and we continue with the description of the SO mechanism and how it is incorporated into the proposed solution, which allows (a) CDN clients (OS) to rationalize their reservation, use and exchange of resources in CDNs and (b) the CDN to also rationalize the reservation and use of resources in meta-CDN, instead of blindly reserving disk space and bandwidth.

Overall, we adopt a Predictive Reservation Scheme (PRS) that involves four, different (operational) aspects (Fig. 2). Such scheme tries to accurately model the load directed by the client (site) and use such information for prompt reservations of CDN resources. To implement these resource reservations within the CDN we adopt the financial instrument of stocks. Although the clients want to predict load as accurately as possible to avoid over- or under-reservation, predictions cannot be 100% accurate. To face failures of the predictive model we introduce the SM, where sites can exchange resources at mutually (between OS) advantageous prices.

Finally, to face failures of the SM, we introduce the SO mechanism, where OS can buy SO at prices and for a duration defined by models such as that of BS. OS can exercise them if and when they are needed (i.e., if an OS runs out of resources at the end of the billing period and cannot find the amount needed through the SM).

### 3.1 The Prediction Mechanism (PM)

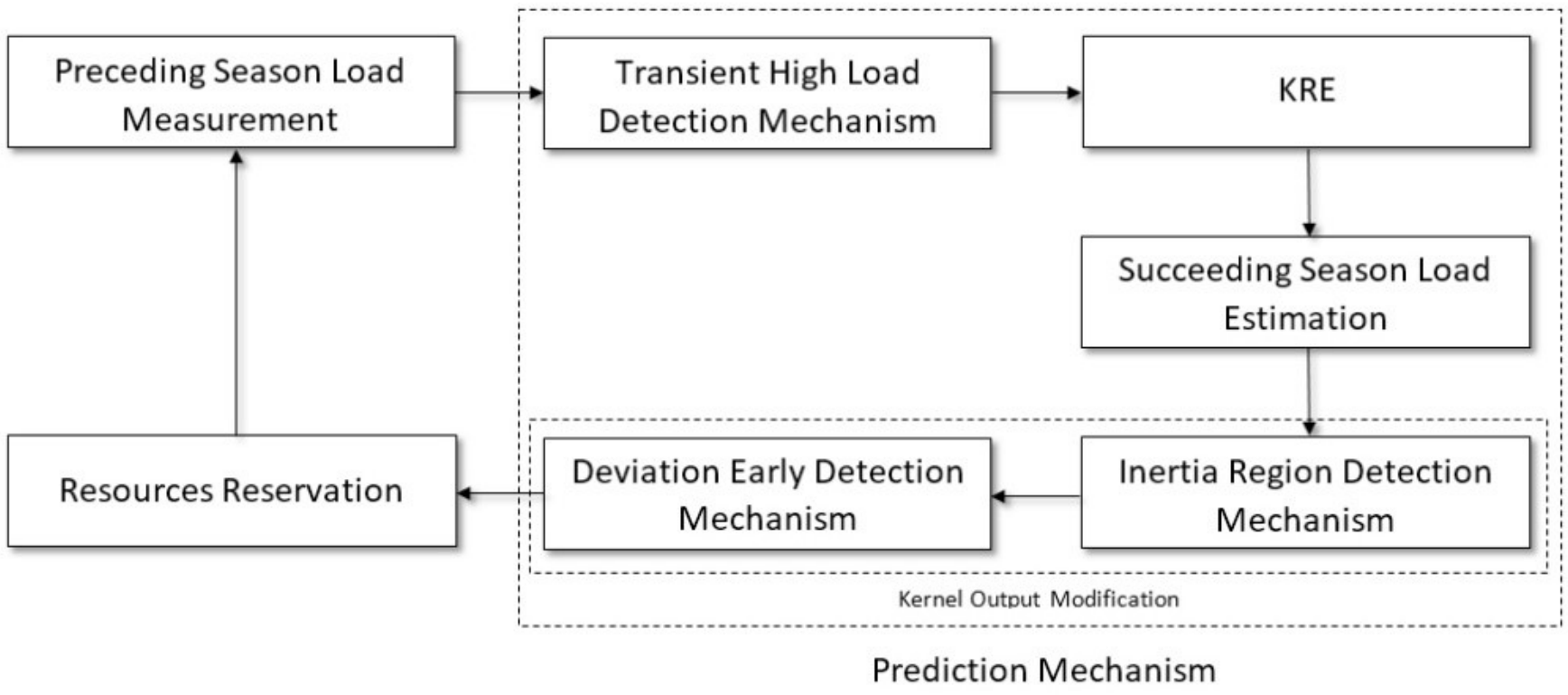

**Fig. 3.** Prediction Mechanism (PM)

The PM itself is the first aspect of PRS [20], [22] and is based on Kernel Regression Estimators (KRE) but also on other complementary techniques, such as:

- **Transient High Load Detection Mechanism** (THLDM), which detects transient high load created by phenomena like Flash Crowds (legitimate load) or DDoS attacks (malicious load) and exclude the corresponding load from the next season forecasting.
- **Kernel Regression Estimators**, which accurately models the site workload, over specific time frames, and outputs a "refined" model of the anticipated load. This refinement involves a stepization of the derived (and constantly updated) function of the load over time.
- **Inertia Region Detection Mechanism** (IRDM), which modifies the KRE output to include negative step-like segments termed load inertia regions, whenever the modeled load leaves local maxima.
- **Deviation Early Detection Mechanism** (DEDM), which monitors with high time granularity the efficiency of the prediction mechanism about the actual load and adjusts the resources reservation for the next time frame (i.e., next day) by modifying the KRE output for that specific time frame.

- **Initial Resource Reservation Monitoring Mechanism** (IRRMM), which applies only to the initial period where no previous actual data exists.

The result of the prediction mechanism is binding as an upper limit regarding the next reservation of resources by the OS. This prevents the system from malicious users (OS) who could reserve many resources in advance at small prices and, subsequently, sell them through the SM mechanism at a certain profit.

### 3.2 The Use of Stocks

To implement the predictive resource reservation within the CDN we adopt the financial instrument of stocks. In the context of our problem, stocks are units of disk space and bandwidth share traded at specific prices. The assumptions that our stock-based modeling relies upon are the following: (a) CDN resources are not infinite i.e., the number of stocks that the CDN can trade at any price is countable and (b) CDN charging for resource reservation/use relies on the volume/quantity and the planned use date/time. Urgent requests for resources are "penalized" with high prices. Reservations confirmed well beforehand are preferentially priced.

The CDN client, OS, buys stocks to immediately reserve resources at current prices available (Fig. 2). When an OS buys stocks, it actually buys the right to use specific resources (disk space or bandwidth) for a specific duration and at a specific price. These rights are easily exchanged via the SM (Fig. 1 and 2). This feature, as well as the overall ease of economic management of resources, are the main reasons why we choose modern capital market tools to design our framework.

### 3.3 The Use of Secondary Market Mechanism

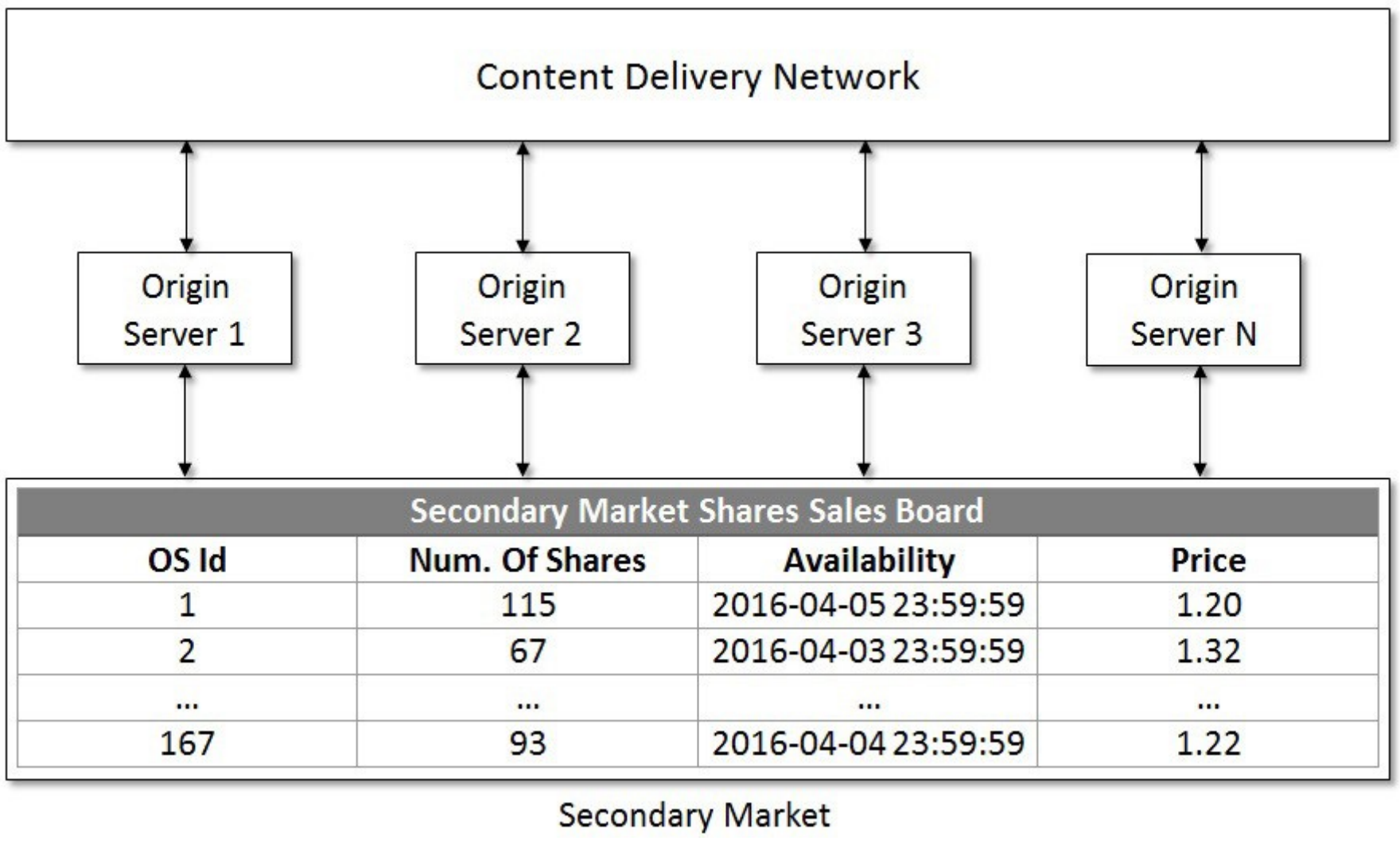

**Fig. 4.** Secondary Market Mechanism

The second line of defense for the failures of the predictive model is the Secondary Market, where neither new financial instruments are issued nor do the issuers, such as corporations, raise new funds. Only investors can purchase from other investors [19].

In our PRS, we establish an SM for stock trading (Fig. 1, 2, and 4). When the reserved resources of an OS are significantly underused, the OS can sell a percentage of the (unused) resources to another OS, through the SM. Thus, the site will reduce, or even eliminate the unused, yet reserved, resources (balancing the prediction failure) and also improve the efficiency on resource use. Conversely, when an OS's reserved resources prove to be insufficient, the OS can seek extra resources through the SMM. Thus, the OS avoids being penalized (overcharged) by the CDN for the needed resources (an implication of prediction failures) and manages to "absorb" the unexpected load through the SM traded resources.

The adoption of the SM proves beneficial for all involved parties (the seller, the buyer and the CDN). The seller reduces its potential loss (or even makes a profit), the buyer adopts a cost-efficient scheme for unforeseen load while the CDN increases its resource utilization indicator, which enables the CDN to offer lower prices and improve its position on competition.

### 3.4 The Use of Stock Options Mechanism

Following the implementation of the SMM, there may still be clients who lack resources at the EOD and who will, therefore, turn to the CDN to buy resources at penalty prices. At this point, we introduce an additional mechanism that follows the philosophy of SOs. This mechanism aims at helping OS with access to resources at non-penalty prices.

A SO is a privilege, sold by one party to another, that gives the buyer the right, but not the obligation, to buy or sell a stock at an agreed-upon price within a specified period. European SOs can only be redeemed at the expiration date. On the other hand, American SOs can be exercised any time between the date of purchase and the expiration date of the option; this greater flexibility for the option buyer results in higher risk for the option seller. As a result, American SOs are more expensive than European SOs.

However, how do we use SOs in our model? In summary, the CDN records how many of its customers, at the EOD, are deficient in resources, how many resources are missing from each and every one of them. Based on this information, forecasts are made for future failures and SOs are purchased, at the beginning of each day, for each of the OSs for future needs. So, each OS pays the cost of acquiring SOs and, when resources are needed, it gets them at a pre-agreed price, simply by exercising SOs he has already bought.

At the end of each day, for each OS, there may be available: a) SOs purchased at the beginning of the day and b) additional SOs that were purchased in the previous days which have not yet been exercised or expired. If the OS actually needs resources, it exercises a corresponding number of SOs and acquires resources at the pre-agreed stocks price.

In our model, we set as the pre-agreed stocks price a) in the case of High Penalty price (Plan 6.a - see Section 4.2.6) the normal price at which CDN sells resources to OS at that time and b) in the case of Low Penalty price (Plan 6.b - see Section 4.2.6) the low penalty price at which CDN sells resources to the OS at that time.

Plan 6.a is the successor to plan 5.a (where the penalty price of the resources is higher than the normal price). In plan 6.a, we introduce the use of SOs. By using SOs, each OS is additionally charged only with the cost of acquiring the SOs, but at the same time gains access to cheaper resources. In detail, the situations that may occur in the EOD for each OS are as follows:

- The resources that can be purchased, based on the remaining active SOs, are more than or equal the resources that need to be purchased by the OS. In this case, the necessary SOs are exercised, and the OS procures the resources it needs at the pre-agreed (normal) price.
- The resources that can be purchased, based on the remaining active SOs, are less than the resources that need to be purchased by the OS. In this case, all available and active SOs are exercised, and the OS procures the corresponding resources at the pre-agreed (normal) price. The remaining resources, for which there are no available SOs, are purchased directly from the CDN at a penalty price.
- No active SOs are available. In this case, all needed resources are purchased directly from the CDN at a penalty price.

Plan 6.b, respectively, is the successor to plan 5.b (where the penalty price of the resources is lower than the normal price). In plan 6.b, we also introduce the use of SO. It is more than obvious that, there are no economic reasons for using SOs in low penalty plan since the additional resources that each OS may need at the EOD can be obtained at a penalty price lower than the standard price (at 95% of the normal price in our simulations).

Why, then, do we also simulate this 6.b plan? The first reason is that we want to confirm our overall model, observe and compare the performance of plan 6.b with the rest of the plans of our model. The second reason is even more critical: Many times, in the markets, on a broader field of applications, SOs are acquired by customers also to ensure that they can access resources in the future. For these cases, therefore, we want to measure the performance of plan 6.b.

Please note that, at the end of each day, ending SOs are deactivated and are no longer available the following day.

### 3.4.1 Pricing of Stock Options

For the pricing of European SOs, tools that are based on closed-form pricing equations of the BS model and their variants, are commonly used. For the pricing of American SOs, many numerical techniques and approximations have been developed, since they do not have closed-form pricing equations. Some of the most popular methods are summarized below.

- **Barone-Adesi & Whaley**. This method [24] separates the value of American options into two parts. The first is the value of a European option, and the second is the value of early exercise. The latter is given by partial differential equations, which Barone-Adesi & Whaley approximate with a quadratic equation (hence the alternative name of the method, the Quadratic Method).
- **Bjerksund & Stensland.** This method was developed in 1993. The method is fast and computationally efficient. For long-dated options, the Bjerksund & Stensland model is more accurate than the Barone-Adesi & Whaley method.
- **Ju & Zhong.** This method [23], first published in 1999, is more accurate than the quadratic approximation for options with short or large maturity times.
- **Binomial and Trinomial Trees.** Both methods involve three general steps: a) A tree for stock prices is constructed. At each time step, the price can either go up or down (for binomial trees). Additionally, trinomial trees allow the stock price to remain the same at each time step. b) The value of the option at maturity is calculated and c) The value of the option at any time before expiry is calculated through backward induction.
  Advantages of these methods are: a) They are easily understood and do not require complex mathematics, b) can be quickly implemented and c) and can be modified to include dividends.
  Disadvantages are: a) they do not produce exact option values (because of their discrete nature) and b) constant volatility is assumed.

### 3.4.2 Pricing Methods of Stock Options that we Evaluate

In the context of our model simulation, we initially examined three methods for calculating SOs costs: the BS (for European SOs, solely for price comparison purposes) as well as the Barone Adesi & Whaley and Binomial and Trinomial Trees methods for American SOs.

The main parameter values we use, in the three methods mentioned above, to calculate the cost of SOs are listed in Table 1, (Simulation Parameters section). One of the essential parameters for pricing SOs is their Time to Maturity (TTM), usually six months, yearly or longer. In our simulations, it turned out that a SO TTM of sixty days was sufficient to almost entirely meet the needs of the OS. In addition, CDN can vary this TTM, depending on the needs.

Figure 5 describes the cost of SO per different method, which corresponds to the normal selling prices of stocks of our model (see also Tables 2 and 3 in section 4.2.6). Correspondingly, Figure 6 describes the cost of SO for the case where the pre-agreed stock prices are equivalent to the lower than normal penalty prices (plan 6.b - section 4.2.6). It should be noted here that the number of sales packages and hence the normal (as also as the low and high penalty) selling prices and their corresponding SO prices are limited. This has as a direct consequence the almost zero computational burden of the CDN by SO mechanism.

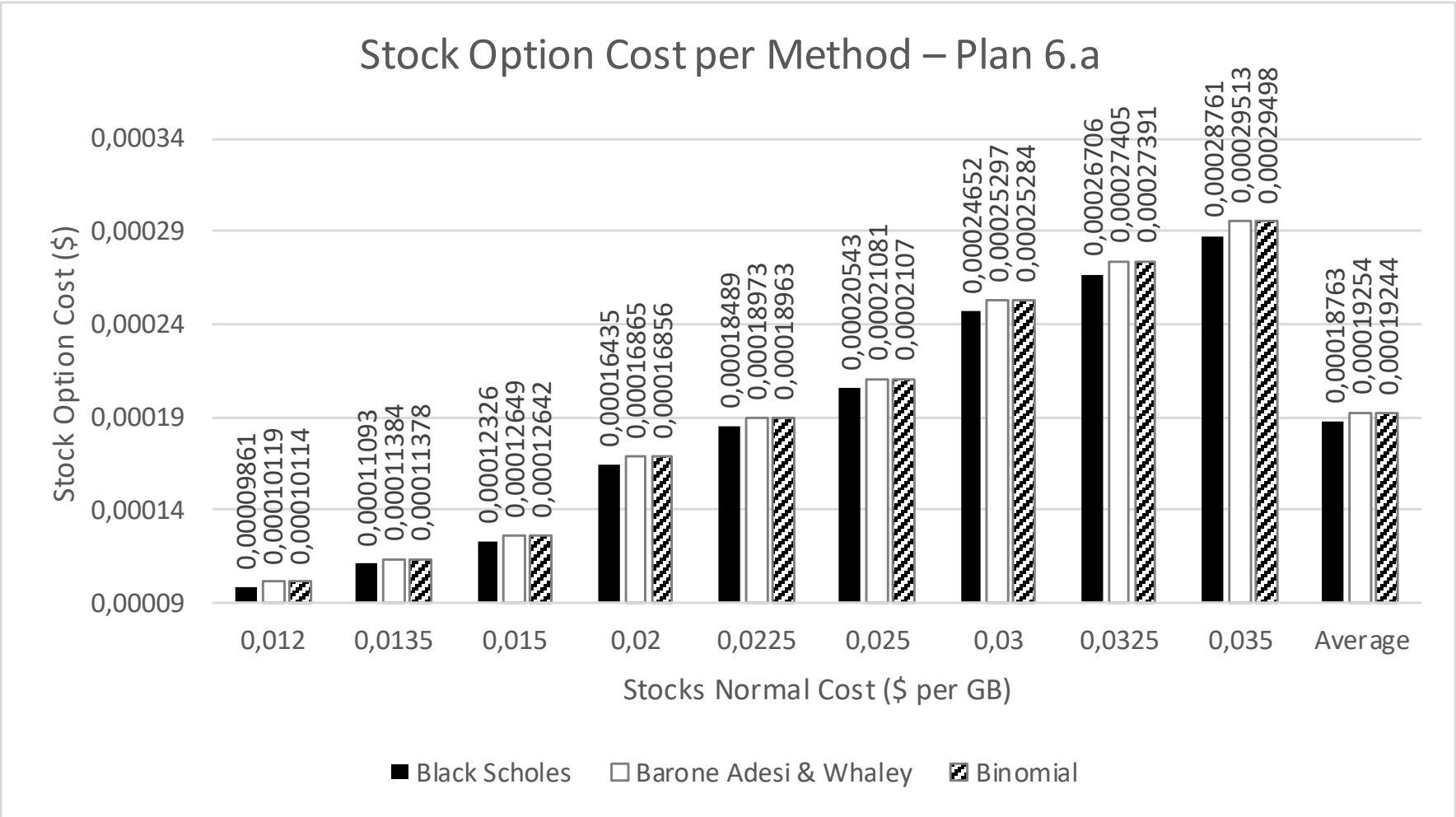

**Fig. 5.** Stock Option Cost per Method – Plan 6.a

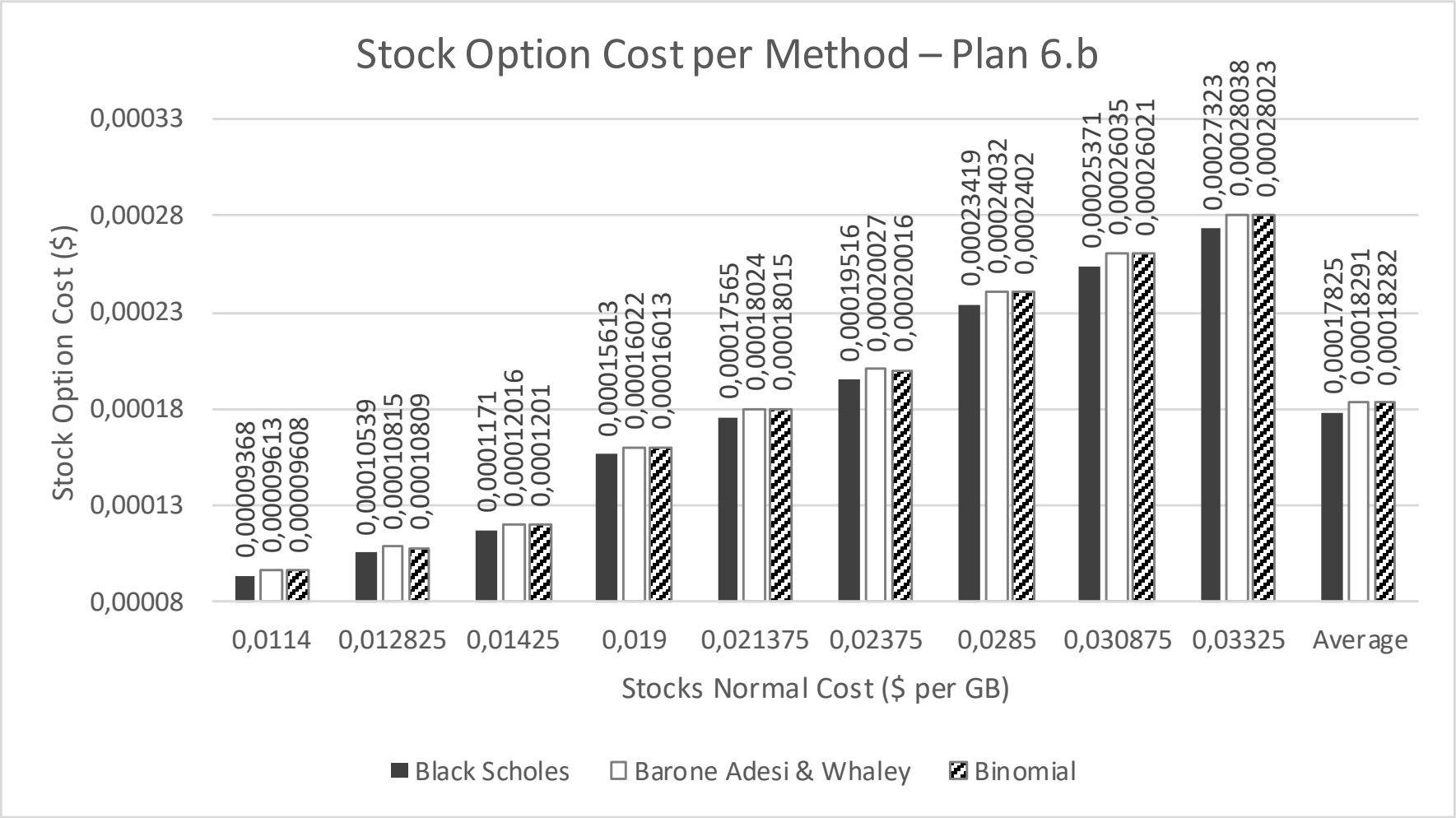

**Fig. 6.** Stock Option Cost per Method – Plan 6.b

SOs calculated using the BS method are the cheapest. SOs calculated by the Barone Adesi & Whaley method are on average 2.62% more expensive than those of BS while SOs calculated by the Binomial method are 2.56% more expensive. In our simulation, we chose to use the Barone Adesi & Whaley method.

### 3.5 Framework Implementation

Our framework can be implemented as software that runs on CDN, evaluate previous season's information, estimates on future workload for each OS and enables the exchange of resources between OS through SM.

# 4. Simulation and Results

In order to assess the performance of the proposed framework, we simulate not only the proposed framework but also the main pricing mechanisms currently available in the market [20,22]. We compare the results and present the effectiveness of our framework with respect to the other solutions currently offered. To compare these plans, we consider that the costs to be reported below include all other CDN charges to OSs, such as Management Fees and Request and Ingress Costs.

The rest of this section is structured as follows: In section 4.1, we describe the basic parameters of the simulations we perform. In section 4.2, we describe the plans that we choose to simulate and compare with our own framework. In Section 4.3.1, we report and compare the simulation results from the OS point of view. In Section 4.3.2, we report and compare the simulation results from the CDN point of view. Finally, in Section 4.4, we comment on the results as a whole.

### 4.1 Simulation Parameters

Our simulations are trace-driven based on cache logs collected from Apache Web Servers on primary Origin Servers (Chambers of Commerce, Universities, Popular Sports Content Sites, etc.). Table 1 summarizes the characteristics of the employed traces as well as the Simulation Parameters.

**Table 1.** List of Simulation Parameters

| Characteristics | Values |
|---|---|
| Origin Servers | 6 |
| Total Period | 52 weeks |
| Reference Period | One week |
| Kernel Function | Gaussian |
| Kernel Bandwidth | Varying from 0.8 to 5.1 |
| Control Time Granularity[3] | 24 hours |
| Min Volume Penalty Calculation | 1 byte |
| Dividend Method (SO) | Continuous |
| Yield Rate (SO) | 0,00 |
| Dividend (SO) | 0,02 |
| Time to maturity (SO) | 60 days |
| Annual risk-free interest rate (SO) | 5,0% |
| Annualized volatility (SO) | 1,0% |

### 4.2 Plans that we Compare

We want to simulate the operation and measure the efficiency of the major plans, which we described in Section 4 of [22] and are currently available on the market. Moreover, we want to compare their efficiency with the efficiency of our own framework. For all the plans that we simulate, we utilize the same number of SCes. The plans we choose to simulate are the following:

[3] Indicates the rate at which measurements are taken, which is, in turn, used for the forecasting task.

### 4.2.1 Pay As You Go Plans

We simulate two PAYG plan variations, the first one with a unified billing model and the second with a staggered one.

### 4.2.2 Pre-Costed Plans

We simulate two variations of Pre-Costed plans: In the first variant (High Penalty), we set the penalty price to 120% of the normal price of each SC. In the second variant (Low Penalty), we set the penalty price to 95% of the normal price of each SC. Moreover, we assume that initially, the seller correctly estimates the Traffic Volume (TV) of the first month and chooses the appropriate SC. Finally, we assume that, for each month that follows, the OS selects the SC that contains, within its TV range, the actual TV that was requested during the previous month.

### 4.2.3 Optimal Cost Pre-Costed Plans

We also simulate two variations of Optimal (Minimum) Cost Pre-Costed plans: one with a high penalty and one with a low penalty. In these simulations, we assume that each OS, during each month, selects the optimal SC that results in the lowest cost. In this way, we want to determine the optimal (lower) cost for each OS, if we assume that every OS can actually guess the optimal SC, for every month, for the actual resources served. Finally, we want to determine the margin for improvement that a CDN has, by properly informing each OS for the selection of the optimum SC, during every billing period.

### 4.2.4 Our Framework

We simulate our own framework and record its functionality and efficiency by activating, in successive stages, the following mechanisms:

a) Forecasting operation without letting the clients (OSs) use the Unused Resources of the day before, without using the SM and SO mechanisms.
b) Forecasting operation along with activating the use of Unused Resources of the day before, but still without using the SM and SO mechanisms.
c) Forecasting operation, use of Unused Resources and SM mechanism, but without using the SO mechanism and
d) Forecasting operation, use of Unused Resources, SM mechanism, and SO mechanism all enabled.

All four variants are simulated with high and low penalties, that is, in total, we simulate eight different variants.

During the first seven days of the simulation, we assume that each OS orders resources that are close to the actual TV that will be experienced. Mechanisms for Unused Resources and SM are available from day 2 and day 1 of the first week, respectively. From the next week (second) the prediction mechanism as also as SO mechanism are activated.

An essential feature of our framework is that pricing is calculated per day rather than per month. In other words, the billing period lasts one day. Correspondingly, the traffic range of the various SCes are calculated per day (Tables 2 and 3). Therefore, assuming that a month lasts 4 weeks rather than 30 days (to facilitate simulations), the P1 SC (in a total of 6 SCes) has a traffic range of 0 GB per day – 0.357 GB per day, the P2 SC 0.357 GB per day to 3.571 GB per day, and so on.

### 4.2.5 Optimal Cost of Our Framework - Ideal Plan (IP)

In this simulation, we calculate the minimum cost that an OS will pay by using our framework. This will happen if the prediction mechanism is 100% accurate or if the SM mechanism succeeds in absorbing all failures of the prediction mechanism. One might say that the Optimal Cost of our framework is equivalent to a unified PAYG plan in which pricing is calculated per day rather than per month. As discussed below and shown in the results, the billing period plays an important role in the cost.

### 4.2.6 Simulation Plans

At this point, we need to mention another important parameter that is included in our simulations. This is the number of SCes available on the market. Initially, 3 SCes were usually offered: low, medium, and high TV. More recently, CDNs began to offer more SCes. We assume that the more SCes are available for an OS to choose from, the more proportional, fair, and low pricing the OS can obtain. Therefore, in our simulations, in all the plans under review, we simulate two sets, with 3 and 6 SCes, respectively, which double the number of simulations.

**Table 2.** Cost and TV Details - 3 SC

| SC | TV | | Normal Cost | High Penalty | Low Penalty | Plan 6.a SO Cost | Plan 6.b SO Cost |
|---|---|---|---|---|---|---|---|
| | (GB per mo.) | (GB per day) | ($ per GB) | ($ per GB) | ($ per GB) | ($ per GB) | ($ per GB) |
| P1 | 0 - 100 | 0 - 3.57 | 0.0325 | 0.039 | 0.030875 | 0.00027405 | 0.00026035 |
| P2 | >100 - 10,000 | >3.57 - 357.14 | 0.0225 | 0.027 | 0.021375 | 0.00018973 | 0.00018024 |
| P3 | >10,000 - 1,000,000 | >357.14 - 35,714.28 | 0.0135 | 0.0162 | 0.012825 | 0.00011384 | 0.00010815 |
| P4 (Unlimited) | >1,000,000 | >35,714.28 | 0.012375 | - | - | - | - |

**Table 3.** Cost and TV Details - 6 SC

| SC | TV | | Normal Cost | High Penalty | Low Penalty | Plan 6.a SO Cost | Plan 6.b SO Cost |
|---|---|---|---|---|---|---|---|
| | (GB per mo.) | (GB per day) | ($ per GB) | ($ per GB) | ($ per GB) | ($ per GB) | ($ per GB) |
| P1 | 0 - 10 | 0 - 0.357 | 0.035 | 0.042 | 0.03325 | 0.00029513 | 0.00028038 |
| P2 | >10 - 100 | >0.357 - 3.57 | 0.030 | 0.036 | 0.0285 | 0.00025297 | 0.00024032 |
| P3 | >100 - 1,000 | >3.57 - 35.71 | 0.025 | 0.03 | 0.02375 | 0.00021081 | 0.00020027 |
| P4 | >1,000 - 10,000 | >35.71 - 357.14 | 0.020 | 0.024 | 0.019 | 0.00016865 | 0.00016022 |
| P5 | >10,000 - 100,000 | >357.14 - 3,571.42 | 0.015 | 0.018 | 0.01425 | 0.00012649 | 0.00012016 |
| P6 | >100,000 - 1,000,000 | >3,571.42 - 35,714.28 | 0.012 | 0.0144 | 0.0114 | 0.00010119 | 0.00009613 |
| P7 (Unlimited) | >1,000,000 | >3,571.42 | 0.011 | - | - | - | - |

The values shown in Tables 2 and 3 are currently used in the market. In Europe and America, prices are usually lower than in other continents. The values of the 3 SCes (Table 2) are the averages of the corresponding values of the 6 SCes (Table 3). For example, in the 6 SCes, P1 has a cost of $0.035 per GB, and a TV range of 0-10 GB while P2 has a cost of $0.030 per GB and a TV range of 10-100 GB. When 3 SCes are available, P1 has a cost of ($0.035 + $0.030) / 2 per GB and a TV range of 0-100 GB. In addition, the SCes P4 and P7 are 9.09% cheaper than the previous ones (P3 and P6) while no penalty is charged.

As indicated in our results below, the number of SCes offered plays a crucial role in the cost that each OS pays, as well as in the utilization rate of reserved resources. Summing up (Table 4), we simulate the following plans:

**Table 4.** Simulated Plans

| Plan | Main Model | SC | Penalty Type | | | | |
|---|---|---|---|---|---|---|---|
| 2.a | Optimal Pre-Costed Plans | 3 & 6 | HP | | | | |
| 2.b | Optimal Pre-Costed Plans | 3 & 6 | LP | | | | |
| 2.c | Pre-Costed Plans | 3 & 6 | HP | | | | |
| 2.d | Pre-Costed Plans | 3 & 6 | LP | | | | |
| Plan | Main Model | SC | Penalty Type | | | | |
| 4.a | PAYG with Staggered Charge | 3 & 6 | n/a | | | | |
| 4.b | PAYG with Unified Charge | 3 & 6 | n/a | | | | |
| Plan | Main Model | SC | Penalty Type | PM | URE | SM | SO |
| 1 | Optimal Proposed Framework | 3 & 6 | n/a | On | On | On | Off |
| 3.1.a | Proposed Framework | 3 & 6 | HP | On | Off | Off | Off |
| 3.1.b | Proposed Framework | 3 & 6 | LP | On | Off | Off | Off |
| 3.2.a | Proposed Framework | 3 & 6 | HP | On | 28 days | Off | Off |
| 3.2.b | Proposed Framework | 3 & 6 | LP | On | 28 days | Off | Off |
| 3.3.a | Proposed Framework | 3 & 6 | HP | On | Full | Off | Off |
| 3.3.b | Proposed Framework | 3 & 6 | LP | On | Full | Off | Off |
| 5.a | Proposed Framework | 3 & 6 | HP | On | Full | On | Off |
| 5.b | Proposed Framework | 3 & 6 | LP | On | Full | On | Off |
| 6.a | Proposed Framework | 3 & 6 | HP | On | Full | On | On |
| 6.b | Proposed Framework | 3 & 6 | LP | On | Full | On | On |

Where HP: High Penalty, LP: Low Penalty, PM: Prediction Mechanism, URE: Unused Resources Exploitation, PAYG: Pay As You Go Plan with Monthly Billing, SO: Stock Options Mechanism, SM: Secondary Market Mechanism

## 4.3 Results

We structure the presentation of the simulation results into two sections, namely the OS results and the CDN results.

### 4.3.1 Results on Origin Servers

#### 4.3.1.1 Actual Traffic Volume Served for each Origin Server

We start the presentation of the OS related results concerning the actual TV served for each OS for the entire simulation period.

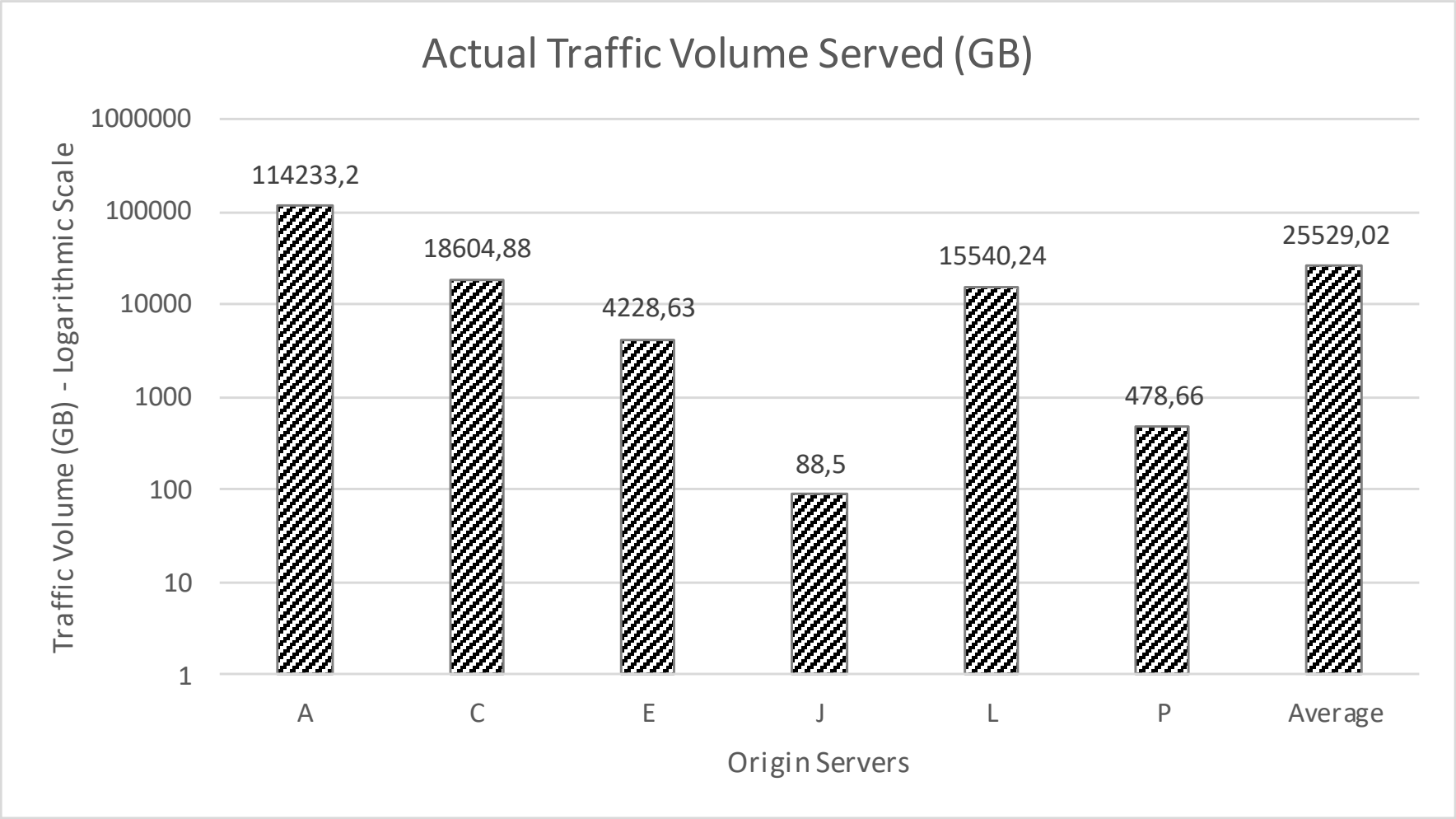

**Fig. 7.** Actual Traffic Volume Served for each Origin Server (GB)

In Figure 7 we see the total Actual TV requested for each of the 6 OSs, which covers a wide range of TV and will help us better study the efficiency and behavior of the plans under consideration.

#### 4.3.1.2 Average Total Cost of Origin Servers

Next, we present the Average Total Cost (Fig. 8) that an OS is required to pay, for the entire simulation period, according to each simulated plan. For the complete results, please see Appendix A.

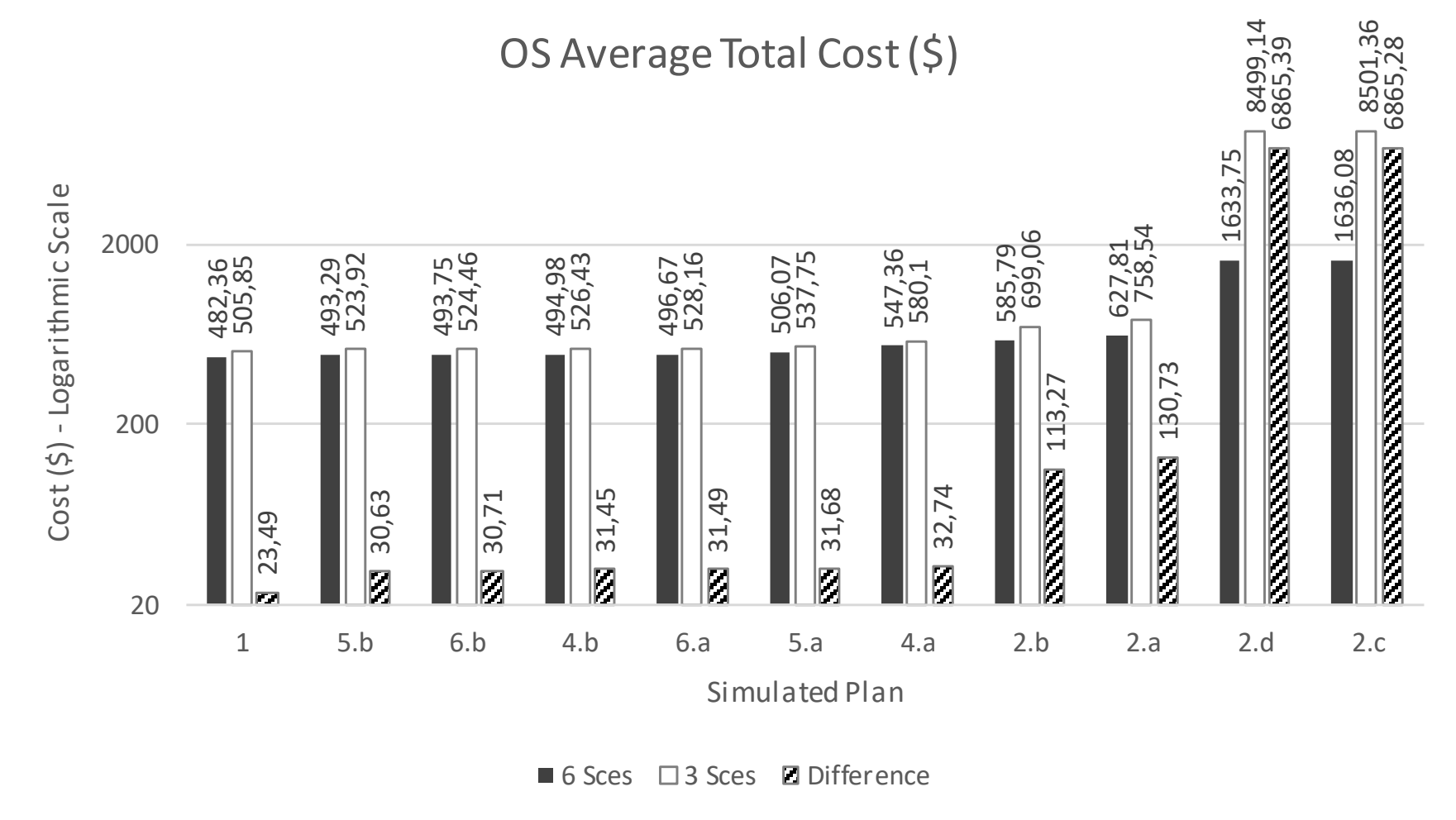

**Fig. 8.** OS Average Total Cost ($)

As we mentioned in [20,22], up to plans 5.a and 5.b (Table 4) and before the introduction of plans 6.a and 6.b that utilize the SO mechanism, plan 5.b had the best performance (Figure 8), achieving the closest (lowest) value to IP (plan 1).

By activating the SO mechanism and referring to the case of 6 SCes and High Penalty plans, the average cost of plan 6.a is only 2.97% more expensive than Ideal plan 1, while the average cost of plan 5.a (before the activation of SO) is 4.91% more expensive. Thus, the activation of the SO mechanism further reduces the average cost of the OS by 1.94%. For the case of 3 SCes, the activation of the SO mechanism further reduces the average cost of the OS by 1,90%.

Respectively, in the Low Penalty plans (5.b and 6.b), the activation of the SO mechanism slightly increases the average cost of the OS by 0.09% (in the case of 6 SCes) and by 0.11% (in the case of 3 SCes). This price increase, essentially stems from the cost of the SO that the OS is required to buy. As we described earlier, we expected this plan (6.b) to be more expensive than 5.b, and we evaluated it mainly for confirmation purposes. Such a plan can be useful to an OS in cases where it needs to secure the availability of resources in the future. Nevertheless, the cost of 6.b is lower than all other plans (except 5.b) under consideration.

Here it is worth noting that the average OS cost of plan 6.b is lower than the average OS cost of plan 6.a. This is because the pre-agreed price for the purchase of resources (in the case of SO exercise) of plan 6.b is 5% lower than the corresponding pre-agreed price of plan 6.a. As a result, at the EOD, the extra resources purchased in the plan 6.b, are 5% cheaper than the ones in plan 6.a.

Summing up, if an **OS selects a model with a lower-than-normal penalty price**, the best (lower cost) plan is 5.b. Plan 6.b is a little bit more expensive (but can be used in case of a requirement to ensure the availability of resources in the future) and PAYG plans 4.b, and 4.a follows. Plan 2.d is much more expensive.

If an **OS selects a model with a higher-than-normal penalty price**, the best (lower cost) plan of our framework is 6.a, and the plan 5.a comes next. PAYG plan 4.b is about 0.34% less expensive plan 6.a while plan 4.a is about 10% more expensive than 6.a. Finally, plan 2.c is much more expensive.

Lastly, the 3 SCes are always more expensive than the 6 SCes, in all plans under consideration.

Overall, the model we propose may, without the need for an estimate of upcoming traffic from the OS, provide the requested services at an OS cost close to the ideal, even lower (plans 5.b and 6.b) than the cost of the PAYG models (which in the actual market is at least twice as high) and clearly lower than the cost of the PCP models. Also, the computational burden for calculating the SO's disposal value is negligible as minimum calculations are required.

#### 4.3.1.3 Average Total Reserved Resources of Origin Servers

In this section, we present the Average Total Resources that OSs reserve and pay (Figure 9) for servicing their needs, according to each plan, for the entire simulation period. For the complete results, please see Appendix B.

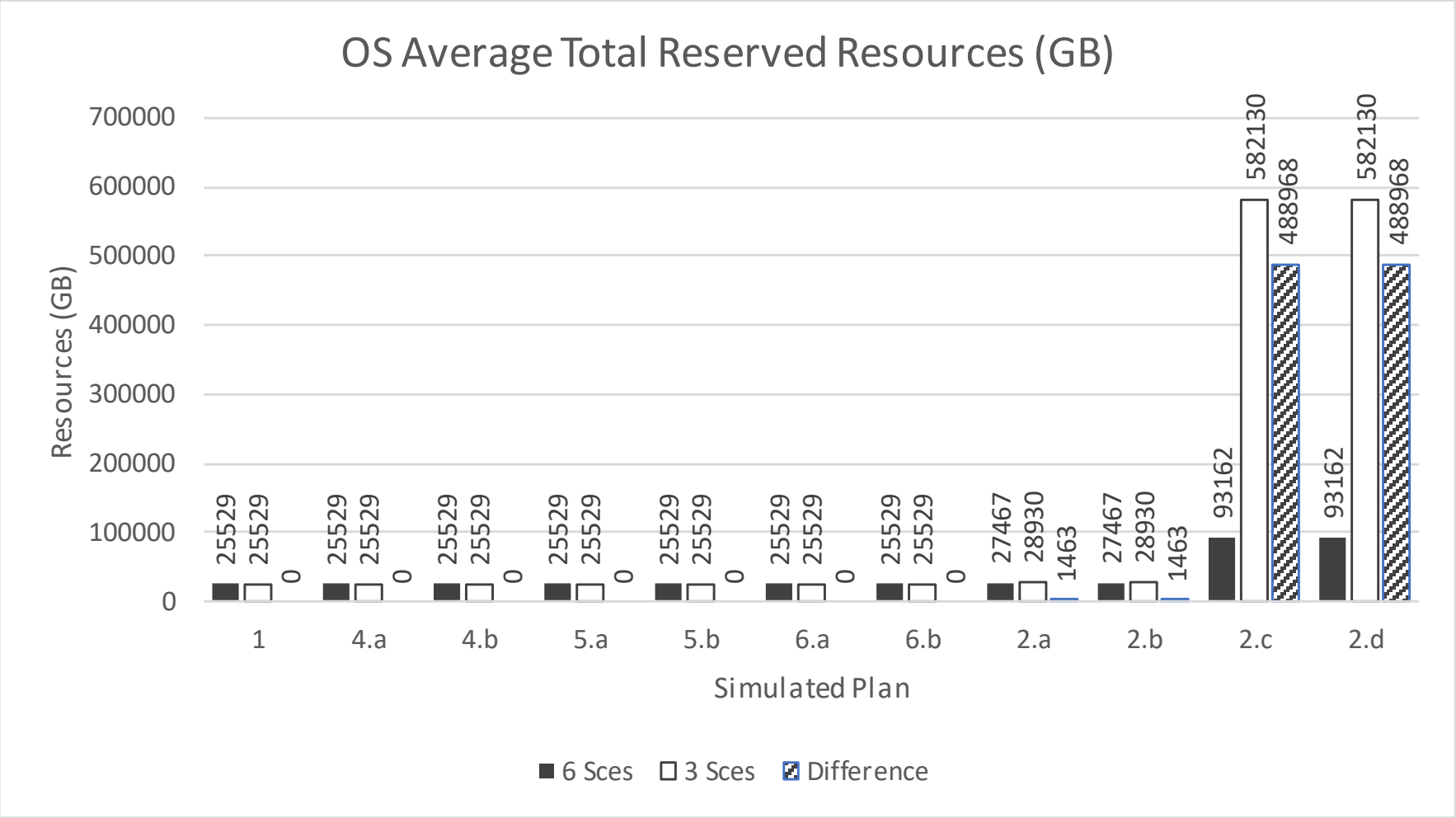

**Fig. 9.** OS Average Total Reserved Resources (GB)

The introduction of the SO mechanism and hence of the plans 6.a and 6.b does not change the image we described at [20,22].

In Pre-Costed Plans 2.c and 2.d, resources reserved and paid, by the OS, are significantly higher than the resources that eventually are required and served to OS customers. It is obvious that with Pre-Costed plans, the OS pays for large amounts of resources that it does not need. On the contrary, in the Optimal PCP 2.a and 2.b, reserved resources are very close to the required resources.

In the case of PAYG plans, 4.a and 4.b, theoretically, no resources are pre-reserved for each OS. OSs consume whatever resources they need and pay only for what they have used. Therefore, we consider that reserved resources of PAYG plans correspond to 100% of the required resources. However, how can a CDN pre-reserve resources and serve the OS without knowing any information about its consumption, or even its consumption prediction? As discussed before, this is probably one of the main reasons that a CDN offers PAYG plans with more than twice the price per GB compared to the Pre-Costed plans.

In our framework (plans 5.a, 5.b, 6.a, and 6.b), reserved resources correspond to 100% of the required resources, regardless of the number of SCes.

In short, the OSs that choose our framework only pay for the resources they need. In addition, and in relation to PAYG plans, CDN has a very good image of future demand and can, therefore, charge at a lower cost per GB and be very competitive with other CDNs that use PAYG plans or Pre-Costed plans.

Lastly, the Penalty type does not affect resources reservation at all in any of the plans considered.

#### 4.3.1.4 Average Cost per Served Resources of Origin Servers

The next metric (that we are considering) is the OS Cost per Served Resources, that is, how much OS pays every actual GB served by CDN (Figure 10), for the entire simulation period. For the complete results, please see Appendix C.

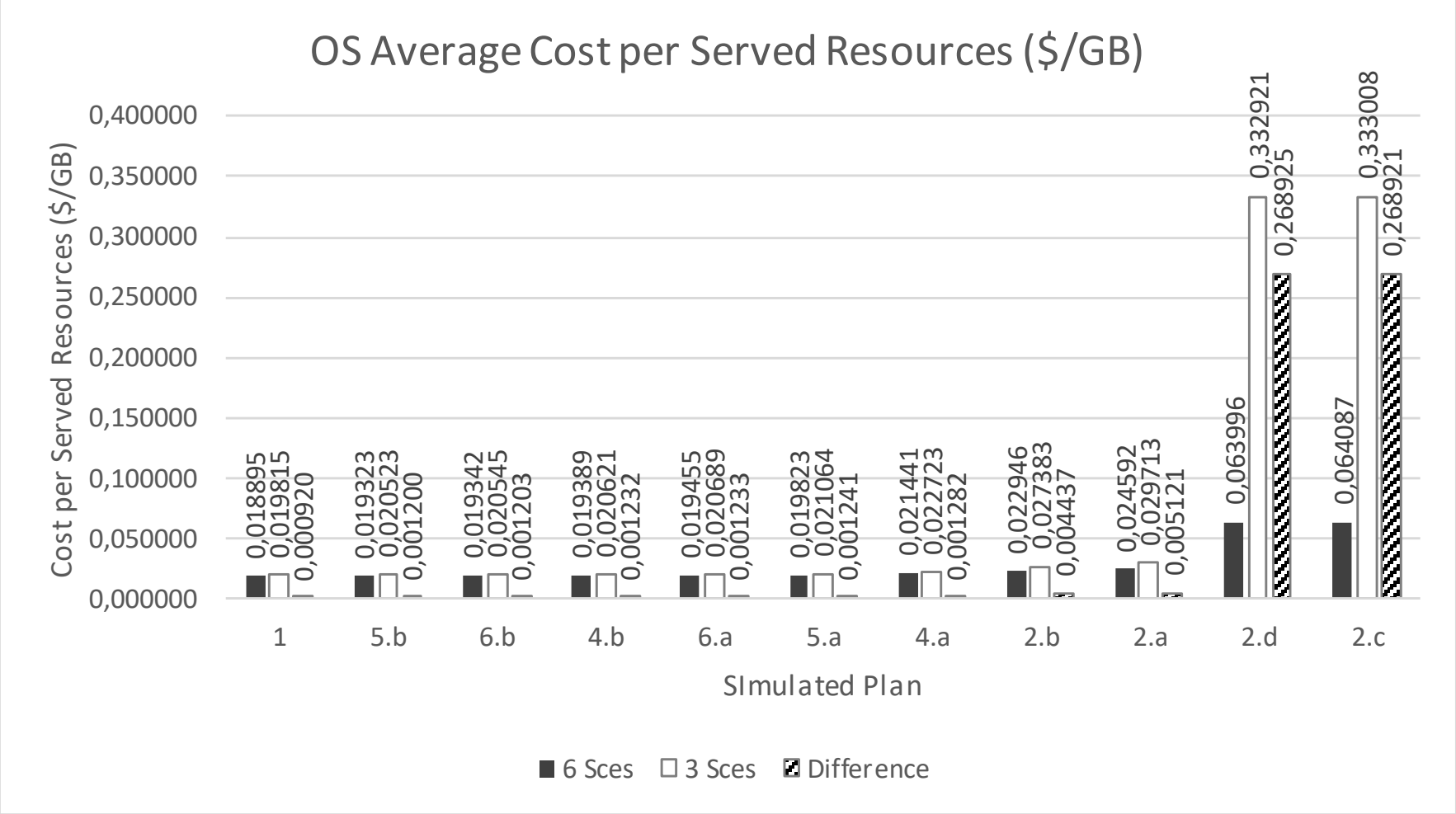

**Fig. 10.** OS Average Cost per Served Resources ($/GB)

In this metric, the worst performance (higher cost per served resources) is expected and confirmed for the Pre-Costed plans 2.c and 2.d (238.70% - 1580.61% more expensive than the IP for 6 and 3 SCes respectively). On the contrary, plans 2.a and 2.b perform much better (21.44% - 49.95% more expensive than the IP for 6 and 3 SCes respectively). In the case of PAYG plans 4.a and 4.b, costs per served resources are significantly reduced (2.61% - 14.68% more expensive than the IP for 6 SCes and 3 SCes respectively).

Our framework (plans 5.b and 5.a) performs even better (2.26% - 6.31% more expensive than the IP for 6 SCes and 3 SCes respectively). Finally, with the engagement of the SO mechanism (plans 6.b and 6.a), the OS cost per served resources ranges from 2.37% - 4.41% more expensive than the IP for 6 SCes and 3 SCes respectively.

Summing up, if an **OS selects a model with a lower-than-normal penalty price**, the best (lower cost per served resources) plan is 5.b. Plan 6.b is a little bit more expensive (but can be used in case of a requirement to ensure the availability of resources in the future), and PAYG plans 4.b, and 4.a follows.

If an **OS selects a model with a higher-than-normal penalty price**, the best (lower cost per served resources) plan is 4.b, 6.a follows being 0.34% more expensive than plan 4.b, the plan 5.a comes next and plan 4.a follows, being about 10% more expensive than 6.a.

Lastly, the 3 SCes are always more expensive (higher cost per served resources) than the 6 SCes, in all plans under consideration.

Overall, the model we propose may, without the need for an estimate of upcoming traffic from the OS, provide the requested services to the OS at a cost per served resources close to the IP, in some cases even lower than the cost per served resources of the PAYG models (which in the actual market is at least twice as high) and clearly lower than the cost per served resources of the PCP models. Also, the computational burden for calculating the SO's disposal value is negligible as minimum calculations are required.

### 4.3.2 Results on Content Delivery Network

In this section, we study the performance of different plans from the point of view of the CDN. Appendix D shows the results of the CDN related Metrics, for the entire simulation period, when 6 SCes are offered, while Appendix E shows similar results when 3 SCes are offered.

Which are the CDN related metrics that are worth examining? CDN cost is primarily due to the purchase of resources from Meta-CDN. This cost also includes the CDN management cost, as mentioned earlier. CDN revenue arises from the sale of resources to the OS. CDN profit results from its revenue minus its expenses. These three metrics are also calculated per served Resources.

In **Pre-Costed plans**, CDN buys the total of the resources requested by the OS for each billing period. In each billing period, if some OSs have surpluses and some deficits, CDNs can first carry out an internal redistribution of resources. If the excess of resources of some OSs is not enough, then and only then, the CDN needs to resort to Meta-CDN to buy additional resources with a penalty price. Any OS that is short of resources is charged with a penalty for its entire deficit and not just for the deficit percentage that was not covered by the internal redistribution.

In **PAYG plans**, we do not know the algorithm that dictates when and how many resources will be purchased from Meta-CDN. They may perform blind Resources reservation from Meta-CDN. Alternatively, assuming that the CDNs have a tracking mechanism for the total resources consumed per billing period, they may reserve for the next billing period as much resources as consumed in the previous one. A third alternative is to reserve an extra percentage of resources in relation to what was consumed in the previous billing period, etc. In any case, not knowing the upcoming load, the CDN is exposed to increased costs that are passed on to the OS. This is confirmed by the cost of PAYG plans in relation to Pre-Costed plans, **in the market**. This is the reason why the prices in the Cost and the Cost per served Resources, of 4.a and 4.b plans, are theoretical, they do not correspond to reality, and we only study them for the theoretical model comparison.

In **our framework**, unlike previous plans, the CDN monitors resources demand. Therefore, using the modeling and forecasting mechanisms as well as the SM and SO mechanisms with which it addresses the forecasting failures, together with the internal redistribution of any surpluses and deficits at the end of the day, significantly reduces the need for extra resources with a penalty price.

#### 4.3.2.1 Total CDN Cost and Total CDN Cost per Served Resources

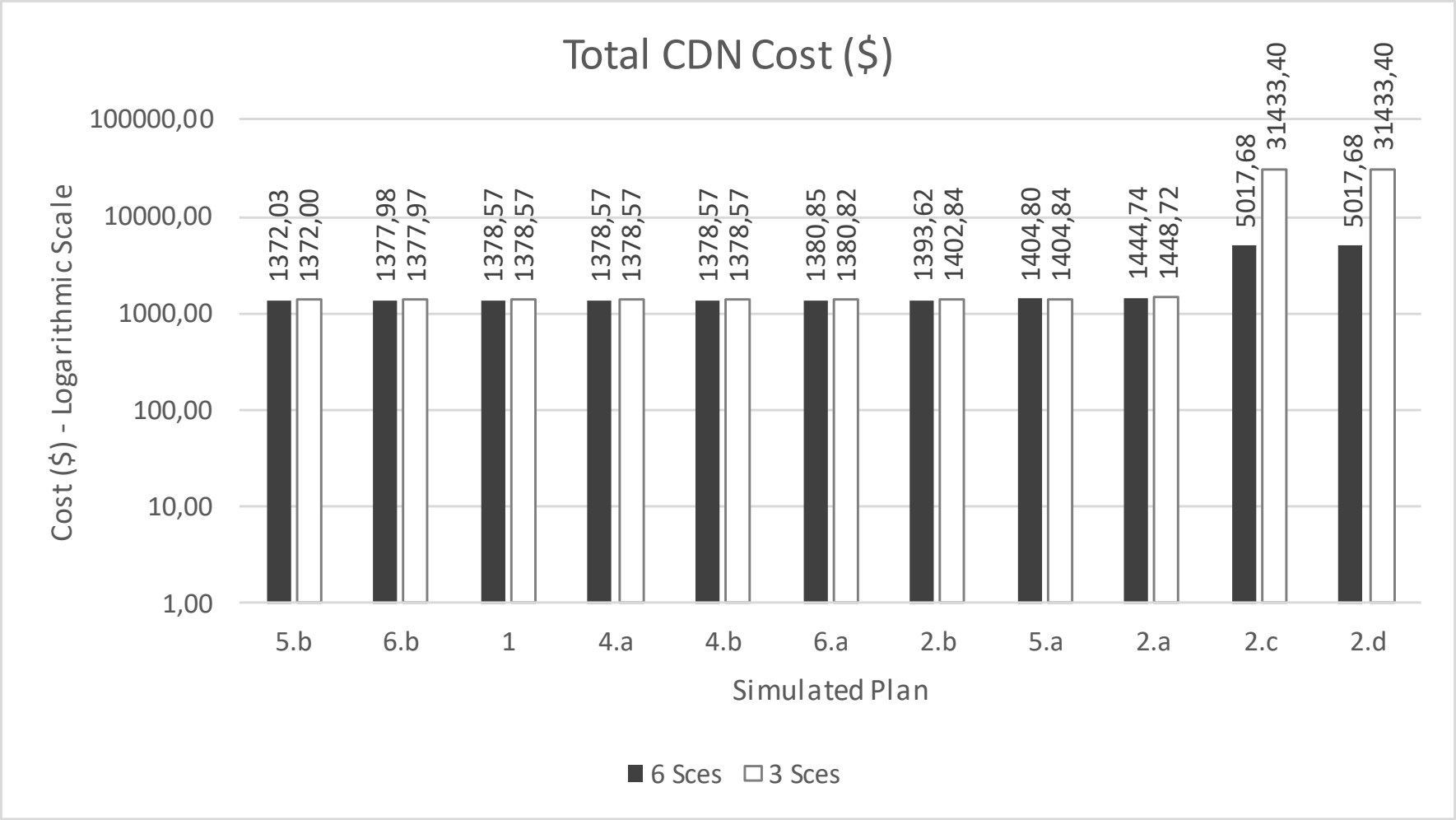

**Fig. 11.** Total CDN Cost ($)

Regarding the metric of Total CDN Cost, as we mentioned in [20,22] up to plans 5.a and 5.b and before the introduction of plans 6.a and 6.b that utilize the SO mechanism, plan 5.b had the best performance (Fig. 11), having a cost even lower than the cost of IP (plan 1), while plan 5.a had 1.8% higher cost than the IP. By activating the SO mechanism, plan 6.a (that corresponds to plan 5.a) is just 0.15% more expensive than the IP, while plan 6.b (that corresponds to plan 5.b) is still less expensive from the IP but more expensive than plan 5.b due to the SO cost.

Moreover, the Total Cost of CDN is very high in Pre-Costed plans 2.c and 2.d while in Optimal Pre-Costed plans, 2.a and 2.b decreases significantly. In PAYG plans 4.a and 4.b, the cost is identical to that of the IP.

However, how can costs of plans 5.b and 6.b be lower than the cost of IP? The primary reasons for this are the following: a) a percentage of the resources is ordered at a penalty price from the CDN, b) the total resources acquired are exactly what is served to the OS and c) the low penalty price (plan b) is lower than the normal price.

Finally, Total CDN Cost is affected by the number of SCes, only in Pre-Costed plans and Optimal Pre-Costed plans. As a conclusion, our framework is performing well and allows CDN to serve OS at a very low cost.

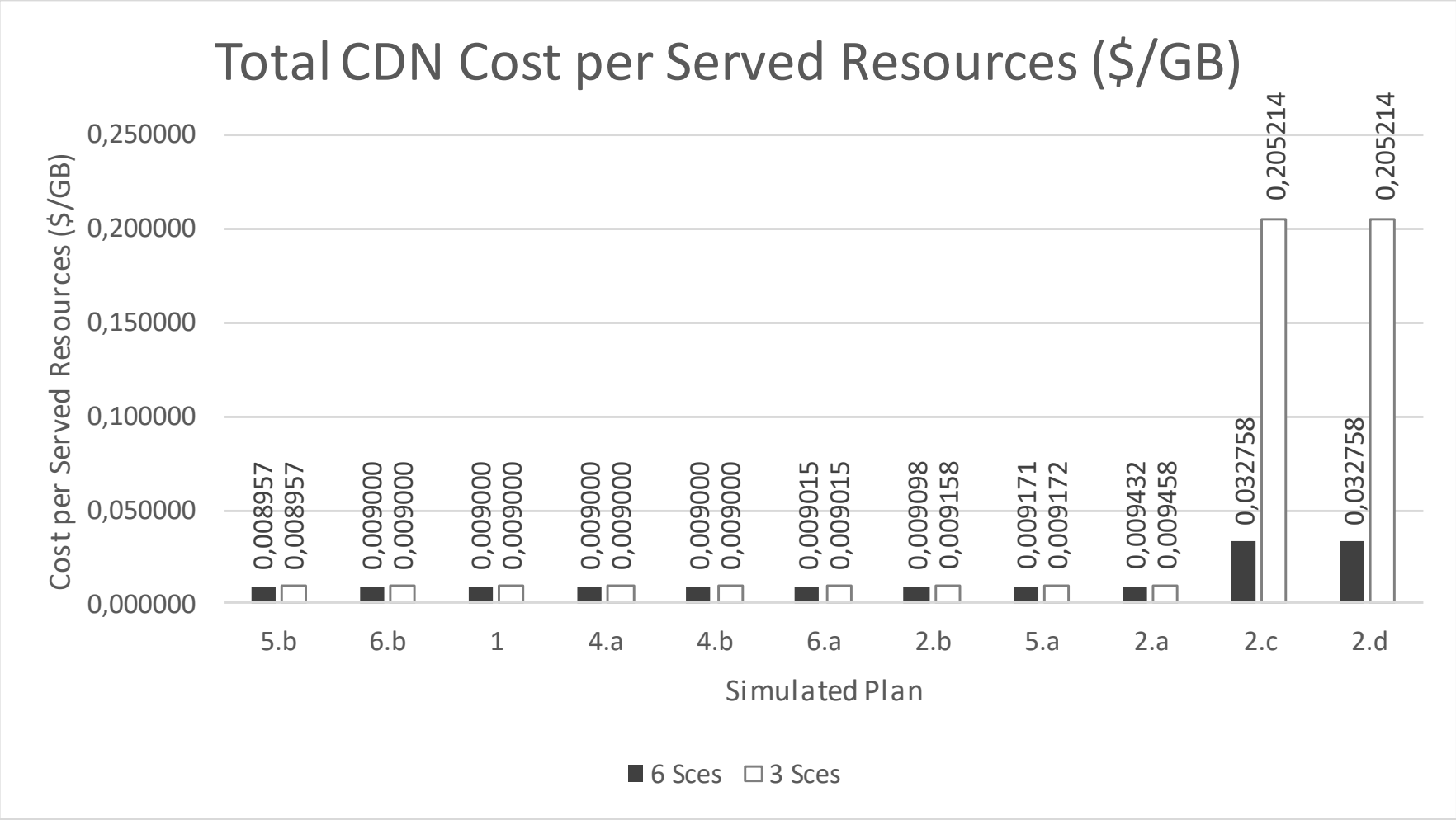

**Fig. 12.** Total CDN Cost per Served Resources ($/GB)

The same observations also apply to the metric of Total CDN Cost per Served Resources (Figure 12).

### 4.3.2.2 Profit to Cost (PtC) Ratio of Content Delivery Networks

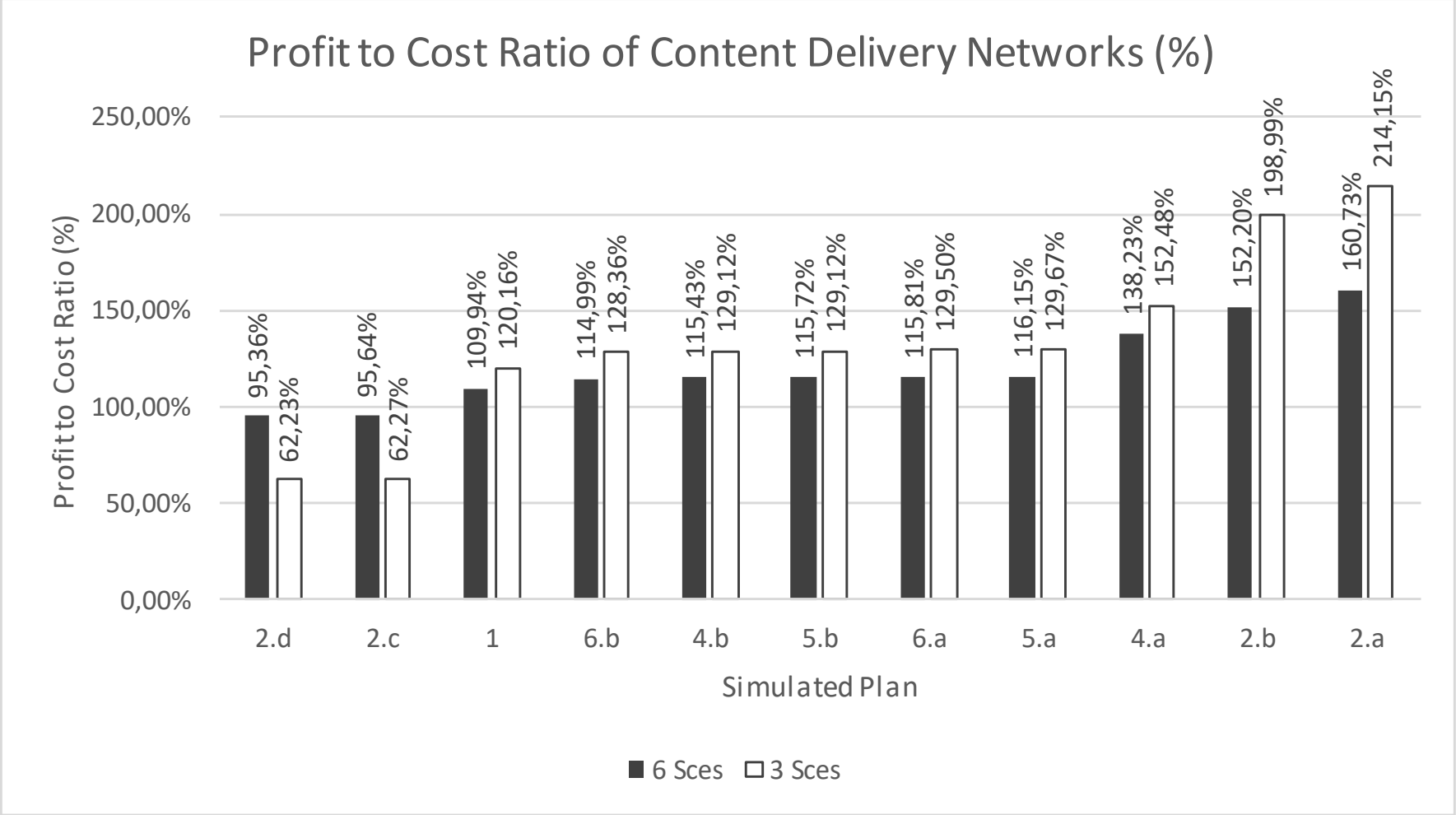

**Fig. 13.** Profit to Cost Ratio of Content Delivery Networks (%)

Figure 13 presents the PtC ratio for each plan under consideration. Evidently, a high ratio indicates a very efficient case.

The lowest PtC ratio is observed in Pre-Costed plans 2.c and 2.d, regardless of the SCes offered. This, coupled with the fact that the cost for the CDN (Figures 11 and 12) as also as the OS cost (Figures 8 and 10) of these two plans is the highest, makes them the worst choice since a CDN has to invest significantly, charge customers substantially and earn moderately.

**Optimal** Pre-Costed plans 2.a and 2.b present the highest PtC ratio with relatively low costs (Figures 11 and 12) for the CDN. However, they are quite expensive for the OS (Figures 8 and 10), so they are not the best choice for them (OS). We also remind that Optimal Pre-Costed plans describe the theoretical optimum of Pre-Costed plans. In order for an OS to approach this optimum, it needs to be adequately informed by the CDN after each billing period as well as to have an accurate estimate of the traffic it will receive in the future.

The results of the **PAYG** plans differ from one another: **plan 4.b** (unified charge) has lower PtC ratio than 2.a and 2.b plans, but it also has low costs for CDN (Figures 11 and 12) and OS (Figures 8 and 10). We, therefore, believe that that is a fair choice for both the CDN and the OS. **Plan 4.a** (staggered charge), compared to plan 4.b, has higher PtC ratio, same cost for the CDN (Figures 11 and 12) but higher OS cost (Figures 8 and 10). This is an important reason (higher OS cost) not to be preferred by the OS and therefore not to be widely offered by the CDN. This is why we consider plan 4.a to be less favorable than plan 4.b.

Regarding our framework, plans 5.a, 6.a, 5.b and 6.b have similar PtC ratios to that of plan 4.b (Figure 13), with PtC ratios of plans 5.a, 6.a and 5.b being slightly better. Taking into account the facts a) that plan 4.b has higher OS costs (Figures 8 and 10) than plans 5.b and 6.b, b) that plan 4.b has higher CDN costs (Figures 11 and 12) than plans 5.b and 6.b and c) that plan 4.b has no information about the upcoming traffic, **we consider our framework the best option overall**. Our scheme serves at the lowest possible cost, sells at the lowest price (except the price of the IP which we are studying for comparison purposes) and its profit to cost ratio is noteworthy, as it exceeds 115% of the CDN cost for 6 SCes and 128% of the CDN cost for 3 SCes.

## 4.4 Discussion

We summarize by answering whether each of the plans under consideration meets the requirements of OS and CDN. The requirements of an OS include the low cost of buying, the optimized use of purchased resources and not being obliged to predict the upcoming traffic. The requirements of a CDN include the low service cost, a reliable forecasting traffic mechanism, the optimized use of purchased resources, a competitive sales price compared to other CDNs, and the highest possible profit.

Pre-Costed plans 2.c and 2.d have the most unfavorable behavior compared to the rest of the plans after pre-committing many resources that, eventually, remain unused. This results in high costs and a relatively smaller percentage profit margin for CDN, but also a high final cost for the OS. In addition, the OS is required to choose a plan before each billing period. We consider them the worst option for CDN and OS.

The theoretical optimal limit of Optimal Pre-Costed plans shows that there is much room for improvement in Pre-Costed plans. A prerequisite for this is that CDNs are to inform the OS after each billing period about the SC that gives them the lower costs, as discussed in [20,22]. If this happens, then the performance of Pre-Costed plans can approach the results of plans 2.a and 2.b, which implies a significant reduction in CDN resource waste and costs, an increase in its profit rate, and a reduction in the final cost to OS.

PAYG plans 4.a and 4.b exhibit even better performance than Optimal Pre-Costed plans, with even lower costs and zero unused resources, since they pay exactly what they use. Specifically, the 4.b plan (unified charge plan) is quite close to the metrics of the IP. However, how many resources are indeed reserved from the CDN, for each OS, before each billing period? The fact that the OS cost of PAYG plans, in the real market, is at least twice as high as that of Pre-Costed plans, reinforces our hypothesis that CDNs do not have efficient prediction mechanisms for the upcoming traffic of each OS. This leads them to increased costs that are passed on to their customers.

On the other hand, the CDNs that insist on offering PAYG plans are likely to benefit from shorter billing periods, e.g., daily, as it happens in IP, that is a PAYG plan with unified charge and daily billing period.

Our framework approaches the theoretical performance of PAYG plans, while plans 5.b and 6.b outperforms it. OS is not required to predict the upcoming traffic or to select a SC for the next billing period. CDN has a reliable predictive and failover mechanism for upcoming traffic per OS, resulting in very high resource utilization. The cost of CDN is low and its profit rate satisfactory. Hence, the final OS cost is also low and, in some cases, even lower than the cost of the IP. We, therefore, believe that our framework is the best choice for both OS and CDN.

Regarding the number of SCes and how it affects the metrics of the plans in question, in Pre-Costed plans 2.c and 2.d, 6 SCes gives better results than 3 SCes in the Unused Resources and Cost fields. In Optimal Pre-costed plans, 2.a and 2.b. the number of SCes has a smaller impact, while it does not affect the rest of the plans.

With regard to the type of penalty, and how it affects the performance of the plans in question, the low penalty gives lower costs in all plans except PCP (2.c and 2.d) due to many reserved resources that remain unused. The penalty type does not affect resource reservation.

# 5. Conclusions and Future Work

In this paper, we present a framework for managing CDN resources, using tools from the capital market. We rely on our previous work that deals with efficient prediction of CDN resource use by OS [20,22] and uses a Secondary Market mechanism to redistribute resources among OS. We enhance our framework by introducing a Stock Options mechanism operation, simulating different plans and ways in which the remaining resources can be redistributed among Origin Servers.

We implemented extensive simulations of all available plans in the market [20,22] as well as our framework. In total, we simulated 34 plans (17 main plans with two sub-plans for 3 and 6 SCes each), covering 6 OSs with a wide range of TV and for 52 weeks.

We compared the simulation results from the point of view of OS and CDN independently. The results show that our framework outperforms all the presented plans. Furthermore, our framework, by improving resource utilization results in the reduction of network congestion.

Apart from our current evaluation of the Stock Options mechanism, our plans for future work include a) the evaluation of the exchange of unused SO through SM mechanism, b) the investigation of the efficiency of other kernel estimators and c) the investigation of the possibility to use more than one estimator in parallel and adopt a mechanism that will automatically choose, between the active estimators, the more efficient one. Finally, we plan to investigate the co-existence of CDNs that use our proposed mechanism with other CDNs that do not, and, also, to evaluate the efficiency of our mechanism for Meta CDNs and Cloud CDNs.

# Appendices

## Appendix A - Average Total Cost of OSs Results

| | 6 SCes | | 3 SCes | | Cost Increase due to number of SCes (From 6 to 3) | Cost Increase due to Low Penalty Use (From Hi to Low) | |
|---|---|---|---|---|---|---|---|
| Plan | $ | % to Plan 1 | $ | % to Plan 1 | % | 6 SCes | 3 SCes |
| 1 | 482.36 | 100.00% | 505.85 | 100.00% | 4.87% | | |
| 2.a | 627.81 | 130.15% | 758.54 | 149.95% | 20.82% | | |
| 2.b | 585.79 | 121.44% | 699.06 | 138.20% | 19.34% | -6.69% | -7.84% |
| 2.c | 1636.08 | 339.18% | 8501.36 | 1680.61% | 419.62% | | |
| 2.d | 1633.75 | 338.70% | 8499.14 | 1680.17% | 420.22% | -0.14% | -0.03% |
| 3.1.a | 576.57 | 119.53% | 610.12 | 120.61% | 5.82% | | |
| 3.1.b | 558.46 | 115.78% | 590.86 | 116.81% | 5.80% | -3.14% | -3.16% |
| 3.2.a | 514.85 | 106.73% | 545.80 | 107.90% | 6.01% | | |
| 3.2.b | 496.35 | 102.90% | 526.32 | 104.05% | 6.04% | -3.59% | -3.57% |
| 3.3.a | 514.96 | 106.76% | 545.84 | 107.91% | 6.00% | | |
| 3.3.b | 496.42 | 102.91% | 526.31 | 104.04% | 6.02% | -3.60% | -3.58% |
| 4.a | 547.36 | 113.48% | 580.10 | 114.68% | 5.98% | | |
| 4.b | 494.98 | 102.62% | 526.43 | 104.07% | 6.35% | | |
| 5.a | 506.07 | 104.91% | 537.75 | 106.31% | 6.26% | | |
| 5.b | 493.29 | 102.27% | 523.92 | 103.57% | 6.21% | -2.53% | -2.57% |
| 6.a | 496.67 | 102.97% | 528.16 | 104.41% | 6.34% | | |
| 6.b | 493.75 | 102.36% | 524.46 | 103.68% | 6.22% | -0.59% | -0.70% |

## Appendix B - Average Total Reserved Resources of OSs Results

| | 6 SCes | | 3 SCes | | Reserved Resources Increase due to number of SCes (From 6 to 3) |
|---|---|---|---|---|---|
| Plan | GB | % to Plan 1 | GB | % to Plan 1 | % |
| 1 | 25529 | 100.00% | 25529 | 100.00% | 0.00% |
| 2.a | 27467 | 107.59% | 28930 | 113.32% | 5.33% |
| 2.b | 27467 | 107.59% | 28930 | 113.32% | 5.33% |
| 2.c | 93162 | 364.93% | 582130 | 2.280.27% | 524.86% |
| 2.d | 93162 | 364.93% | 582130 | 2.280.27% | 524.86% |
| 3.1.a | 29091 | 113.96% | 29091 | 113.96% | 0.00% |
| 3.1.b | 29091 | 113.96% | 29091 | 113.96% | 0.00% |
| 3.2.a | 25529 | 100.00% | 25529 | 100.00% | 0.00% |
| 3.2.b | 25529 | 100.00% | 25529 | 100.00% | 0.00% |
| 3.3.a | 25529 | 100.00% | 25529 | 100.00% | 0.00% |
| 3.3.b | 25529 | 100.00% | 25529 | 100.00% | 0.00% |
| 4.a | 25529 | 100.00% | 25529 | 100.00% | 0.00% |
| 4.b | 25529 | 100.00% | 25529 | 100.00% | 0.00% |
| 5.a | 25529 | 100.00% | 25529 | 100.00% | 0.00% |
| 5.b | 25529 | 100.00% | 25529 | 100.00% | 0.00% |
| 6.a | 25529 | 100.00% | 25529 | 100.00% | 0.00% |
| 6.b | 25529 | 100.00% | 25529 | 100.00% | 0.00% |

## Appendix C - Average Cost per Served Resources of OSs Results

| - | 6 SCes | | 3 SCes | | "Cost Increase due to number of SCes (From 6 to 3)" | "Cost Increase due to Low Penalty Use (From Hi to Low)" | |
|---|---|---|---|---|---|---|---|
| Plan | ($/GB)" | % to Plan 1 | ($/GB)" | % to Plan 1 | % | 6 SCes | 3 SCes |
| 1 | 0.018895 | 100.00% | 0.019815 | 100.00% | 4.87% | | |
| 2.a | 0.024592 | 130.15% | 0.029713 | 149.95% | 20.82% | | |
| 2.b | 0.022946 | 121.44% | 0.027383 | 138.19% | 19.34% | -6.69% | -7.84% |
| 2.c | 0.064087 | 339.18% | 0.333008 | 1680.61% | 419.62% | | |
| 2.d | 0.063996 | 338.70% | 0.332921 | 1680.17% | 420.22% | -0.14% | -0.03% |
| 3.1.a | 0.022585 | 119.53% | 0.023899 | 120.61% | 5.82% | | |
| 3.1.b | 0.021875 | 115.78% | 0.023145 | 116.81% | 5.80% | -3.14% | -3.16% |
| 3.2.a | 0.020167 | 106.73% | 0.021379 | 107.90% | 6.01% | | |
| 3.2.b | 0.019443 | 102.90% | 0.020617 | 104.05% | 6.04% | -3.59% | -3.57% |
| 3.3.a | 0.020172 | 106.76% | 0.021381 | 107.91% | 6.00% | | |
| 3.3.b | 0.019445 | 102.91% | 0.020616 | 104.04% | 6.02% | -3.60% | -3.58% |
| 4.a | 0.021441 | 113.47% | 0.022723 | 114.68% | 5.98% | | |
| 4.b | 0.019389 | 102.61% | 0.020621 | 104.07% | 6.36% | | |
| 5.a | 0.019823 | 104.91% | 0.021064 | 106.31% | 6.26% | | |
| 5.b | 0.019323 | 102.26% | 0.020523 | 103.57% | 6.21% | -2.53% | -2.57% |
| 6.a | 0.019455 | 102.97% | 0.020689 | 104.41% | 6.34% | | |
| 6.b | 0.019341 | 102.37% | 0.020544 | 103.68% | 6.22% | -0.58% | -0.70% |

## Appendix D - CDN Metrics Results (6 SCes)

| Plan | Resources Served (GB) | Resources Reserved (GB) | Revenue ($) | Revenue per Served Resources ($/GB) | Cost ($) | Cost per Served Resources ($/GB) | Profit ($) | Profit per Served Resources ($/GB) |
|---|---|---|---|---|---|---|---|---|
| 1 | 153174 | 153174 | 2894.18 | 0.018895 | 1378.57 | 0.009000 | 1515.61 | 0.009895 |
| 2.c | 153174 | 558977 | 9816.47 | 0.064087 | 5017.68 | 0.032758 | 4798.79 | 0.031329 |
| 2.d | 153174 | 558977 | 9802.53 | 0.063996 | 5017.68 | 0.032758 | 4784.85 | 0.031238 |
| 2.a | 153174 | 164802 | 3766.88 | 0.024592 | 1444.74 | 0.009432 | 2322.13 | 0.015160 |
| 2.b | 153174 | 164802 | 3514.72 | 0.022946 | 1393.62 | 0.009098 | 2121.11 | 0.013848 |
| 4.a | 153174 | 153174 | 3284.16 | 0.021441 | 1378.57 | 0.009000 | 1905.59 | 0.012441 |
| 4.b | 153174 | 153174 | 2969.86 | 0.019389 | 1378.57 | 0.009000 | 1591.29 | 0.010389 |
| 5.a | 153174 | 153174 | 3036.40 | 0.019823 | 1404.80 | 0.009171 | 1631.60 | 0.010652 |
| 5.b | 153174 | 153174 | 2959.72 | 0.019323 | 1372.03 | 0.008957 | 1587.69 | 0.010365 |
| 6.a | 153174 | 153174 | 2980.05 | 0.019455 | 1380.85 | 0.009015 | 1599.20 | 0.010440 |
| 6.b | 153174 | 153174 | 2962.52 | 0.019341 | 1377.98 | 0.008996 | 1584.54 | 0.010345 |

## Appendix E - CDN Metrics Results (3 SCes)

| Plan | Resources Served (GB) | Resources Reserved (GB) | Revenue ($) | Revenue per Served Resources ($/GB) | Cost ($) | Cost per Served Resources ($/GB) | Profit ($) | Profit per Served Resources ($/GB) |
|---|---|---|---|---|---|---|---|---|
| 1 | 153174 | 153174 | 3035.10 | 0.019815 | 1378.57 | 0.009000 | 1656.53 | 0.010815 |
| 2.c | 153174 | 3492781 | 51008.17 | 0.333008 | 31433.40 | 0.205214 | 19574.77 | 0.127794 |
| 2.d | 153174 | 3492781 | 50994.85 | 0.332921 | 31433.40 | 0.205214 | 19561.45 | 0.127707 |
| 2.a | 153174 | 173584 | 4551.21 | 0.029713 | 1448.72 | 0.009458 | 3102.50 | 0.020255 |
| 2.b | 153174 | 173584 | 4194.35 | 0.027383 | 1402.84 | 0.009158 | 2791.51 | 0.018224 |
| 4.a | 153174 | 153174 | 3480.59 | 0.022723 | 1378.57 | 0.009000 | 2102.03 | 0.013723 |
| 4.b | 153174 | 153174 | 3158.59 | 0.020621 | 1378.57 | 0.009000 | 1780.03 | 0.011621 |
| 5.a | 153174 | 153174 | 3226.53 | 0.021064 | 1404.84 | 0.009172 | 1821.69 | 0.011893 |
| 5.b | 153174 | 153174 | 3143.53 | 0.020523 | 1372.00 | 0.008957 | 1771.53 | 0.011565 |
| 6.a | 153174 | 153174 | 3168.95 | 0.020689 | 1380.82 | 0.009015 | 1788.13 | 0.011674 |
| 6.b | 153174 | 153175 | 3146.74 | 0.020544 | 1377.97 | 0.008996 | 1768.78 | 0.011547 |